\RequirePackage{fix-cm}
\documentclass[smallextended]{svjour3}       
\smartqed  
\usepackage[utf8]{inputenc}

\PassOptionsToPackage{hyphens}{url}\usepackage{hyperref}

\usepackage{csquotes}

\usepackage{graphicx}
\graphicspath{{images/}}

\usepackage{multirow}

\usepackage{makecell}

\usepackage{array}

\usepackage{float}

\usepackage{tcolorbox}
\definecolor{amber}{rgb}{1.0, 0.75, 0.0}
\newtcolorbox{mybox}{colback=amber!10,colframe=amber}

\usepackage{natbib}

\usepackage{eurosym}

\usepackage{amssymb}

\usepackage{tcolorbox}

\journalname{Empirical Software Engineering}
\begin{document}

\title{Do RESTful API Design Rules Have an Impact on the Understandability of Web APIs?}

\subtitle{A Web-Based Experiment with API Descriptions}

\titlerunning{Do Design Rules Have an Impact on the Understandability of Web APIs?}

\author{
    Justus Bogner \and
    Sebastian Kotstein \and
    Timo Pfaff
}

\authorrunning{Bogner, Kotstein, Pfaff} 

\institute{
    Justus Bogner \at
        University of Stuttgart, Institute of Software Engineering, Germany\\
        \email{justus.bogner@iste.uni-stuttgart.de}
    \and
    Sebastian Kotstein \at
        Reutlingen University, Herman Hollerith Zentrum, Germany\\
        \email{sebastian.kotstein@reutlingen-university.de}
    \and
    Timo Pfaff \at
        Independent researcher, Germany\\
        Research conducted while at\\
        University of Stuttgart, Institute of Software Engineering, Germany\\
}

\date{Received: date / Accepted: date}

\maketitle

\begin{abstract}

\textit{Context:}
Web APIs are one of the most used ways to expose application functionality on the Web, and their understandability is important for efficiently using the provided resources.
While many API design rules exist, empirical evidence for the effectiveness of most rules is lacking.

\textit{Objective:}
We therefore wanted to study 1) the impact of RESTful API design rules on understandability, 2) if rule violations are also perceived as more difficult to understand, and 3) if demographic attributes like REST-related experience have an influence on this.

\textit{Method:}
We conducted a controlled Web-based experiment with 105 participants, from both industry and academia and with different levels of experience.
Based on a hybrid between a \textit{crossover} and a \textit{between-subjects} design, we studied 12 design rules using API snippets in two complementary versions: one that adhered to a \textit{rule} and one that was a \textit{violation} of this rule.
Participants answered comprehension questions and rated the perceived difficulty.

\textit{Results:}
For 11 of the 12 rules, we found that \textit{violation} performed significantly worse than \textit{rule} for the comprehension tasks.
Regarding the subjective ratings, we found significant differences for 9 of the 12 rules, meaning that most violations were subjectively rated as more difficult to understand.
Demographics played no role in the comprehension performance for \textit{violation}.

\textit{Conclusions:}
Our results provide first empirical evidence for the importance of following design rules to improve the understandability of Web APIs, which is important for researchers, practitioners, and educators.

\end{abstract}

\section*{Declarations}

\noindent Funding: not applicable\\
\noindent Conflicts of interest: not applicable\\
\noindent Availability of data and material: \url{https://doi.org/10.5281/zenodo.7381500}\\
\noindent Code availability: not applicable\\

\section{Introduction}
Technologies for Web Application Programming Interfaces (APIs) like WSDL, SOAP, and HTTP are the technical foundations for realizing modern Web applications~\citep{Jacobson2011}. 
These technologies allow developers to independently implement smaller software components and share their functionality on the Internet. 
Successfully reusing an existing component, however, depends on the ability to understand its purpose and behavior, especially its API, which hides internal logic and complexity. 
In cases where neither additional documentation nor developers of the original component can be consulted, the Web API might even be the first and only point of contact with the exposed functionality. 
Therefore, understandability is an important quality attribute for the design of Web APIs~\citep{article:Palma2017}.

Over the last two decades, HTTP combined with other well-established Web standards like URI has become a popular choice for realizing Web APIs that expose their functionality through Web resources~\citep{Schermann2016,Bogner2019}.
In these resource-oriented Web APIs, the role of HTTP has shifted from a transport mechanism for XML-based messages to an application-layer protocol for interacting with the respective API~\citep{conference:Pautasso2008}.
With Representational State Transfer (REST)~\citep{article:PrincipleDesignOfTheModernWebArchitecture:2002}, there exists an architectural style that formalizes the proper use of Web technologies like HTTP and URIs in Web applications.
REST is considered a foundation for high-quality, so-called \textit{RESTful} API design, and it describes a set of constraints for the recommended behavior of Web applications, e.g, HTTP-based Web APIs.
However, it does not instruct developers how to implement this behavior~\citep{article:Rodriguez2016}.

Since there exist various interpretations and (mis-)understandings among practitioners how RESTful API design looks like, users and integrators of these services are confronted with a multitude of heterogeneous interface designs, which can make it difficult to understand a given Web API~\citep{conference:Palma2021}.
Therefore, several works have proposed design rules and best practices to complement the original REST constraints and to guide developers when designing and implementing Web APIs, e.g., \citet{Pautasso2014}, \citet{article:Palma2017}, \citet{Richardson2007}, and \citet{Masse2011}.
In most cases, however, we do not have sufficient empirical evidence for the effectiveness of these RESTful API design rules, i.e., if they really have a positive impact on the quality of Web APIs.

Additionally, multiple studies analyzed the degree of REST compliance in practice by systematically comparing real-world Web APIs against proposed design rules. 
Many of these works, e.g., \citet{Neumann2018}, \citet{conference:Renzel2012}, and \citet{article:Rodriguez2016}, concluded that only a small degree of real-world Web APIs are truly RESTful. 
This suggests that many practitioners perceive proposed design rules differently in terms of their importance. 
We have provided confirmation for this in previous work~\citep{Kotstein2021}.
In a Delphi study, we confronted industry practitioners with 82 RESTful API design rules by \citet{Masse2011} to find out which ones they perceived as important and how they perceived their impact on software quality. 
Only 45 out of 82 rules were rated with high or medium importance, and maintainability and usability were the most associated quality attributes.
Both of these attributes are closely related to understandability.

To confirm these opinion-based results with additional empirical evidence, we conducted a controlled Web-based experiment, in which we presented 12 Web API snippets to 105 participants with at least basic REST-related experience.
Each API snippet existed in two versions, one adhering to a design rule and one violating the rule (see, e.g., Fig.~\ref{fig:questionTypeReturn}).
The participants' task was to answer comprehension questions about each snippet, while we measured the required time.
Furthermore, participants also had to rate the perceived difficulty to understand an API snippet. 
In this paper, we present the design and results of our controlled experiment on the understandability impact of RESTful API design rules. 

\section{Background and Related Work}
We start with a discussion of terminology around Web APIs and REST and explain how we use these terms in the paper.
Furthermore, we mention existing works that propose rules and best practices for RESTful API design, and present existing studies about Web API quality.

\subsection{Terminology}
In this paper, we focus on resource-oriented HTTP-based Web APIs.
In distinction to SOAP-/WSDL-based APIs~\citep{conference:Pautasso2008}, we use the term \textit{Web API} for any resource-oriented API that exposes its functionality via HTTP and URIs at the application level.
We consider a Web API as \textit{RESTful}, i.e., a so-called \textit{RESTful API}, if the respective Web API satisfies all mandatory REST constraints defined by~\cite{article:PrincipleDesignOfTheModernWebArchitecture:2002}.
As a consequence, a Web API that implements only some RESTful API design rules cannot automatically be considered as RESTful.

\subsection{Best Practices for REST in Practice}
To combat the potentially harmful heterogeneity of interface designs among Web APIs, several works tried to translate the REST constraints into more concrete guidelines to instruct developers how to achieve good RESTful design with HTTP.
RESTful API design rules and best practices have been proposed in scientific articles, e.g., by~\citet{Pautasso2014},~\citet{conference:Petrillo2016},~\citet{article:DetectionOfRESTPatternsAndAntipatterns}, and~\citet{article:Palma2017}, but also in textbooks, e.g., by~\citet{Richardson2007},~\citet{Masse2011}, and~\cite{book:RestInPractice:2010}.
Moreover, Leonard Richardson developed a maturity model~\citep{web:MaturityModel} allowing Web developers to estimate the degree of REST compliance of their Web APIs.
Incorporating the principles of REST, the model defines four levels of maturity.

\begin{itemize}
    \item Level 0: Web APIs offer their functionality over a single URI and use HTTP solely as a transport protocol for tunneling requests through this endpoint by using \texttt{POST}. Examples of level 0 are SOAP- and XML-RPC-based services.
    \item Level 1: Web APIs use the concept of resources, i.e., expose different URIs for different resources. However, operations are still identified via URIs or specified in the request payload rather than by different HTTP methods.
    \item Level 2: Web APIs use HTTP mechanisms and semantics, including different HTTP methods for different operations and semantically correct status codes. Level 2 partially aligns with the \textit{Uniform Interface} constraint.
    \item Level 3: Web APIs additionally conform to the HATEOAS constraint by embedding hypermedia controls into responses to advertise semantic relations between resources and to offer navigational support to clients.
\end{itemize}

Despite the existence of these design rules and guidelines, multiple studies, e.g., by \citet{Neumann2018}, \citet{conference:Renzel2012}, and \citet{article:Rodriguez2016}, revealed that only a small number of existing Web APIs are indeed RESTful, although many Web APIs claimed to be RESTful~\citep{Neumann2018}. 
This suggests that there is still no common understanding of how a RESTful API should look like in industrial practice. 
Providing empirical evidence may therefore help to identify the effective and important rules from the large collection of existing guidelines.
 
\subsection{Related Work}
Several works investigated the quality of real-world Web APIs, mainly by analyzing interfaces, their descriptions, and exchanged HTTP messages.

\citet{article:Rodriguez2016} analyzed more than 78 GB of HTTP traffic to gain insights into the use of best practices in Web API design. 
They applied API-specific heuristics to extract API-related messages from the whole data set and validated these extracted API requests and responses against 18 heuristics aligned with REST design principles and best practices. 
For a few heuristics, they described their negative effect on maintainability and evolvability when violating associated design principles and best practices. 
Moreover, they mapped the heuristics to the levels of the Richardson maturity model to estimate the level of REST compliance of investigated Web APIs.
The paper concluded that only a few APIs reach level 3 of the maturity model, but the majority of investigated APIs complied with level 2.

The governance of RESTful APIs was the focus for \citet{Haupt2018}: using a framework developed by~\citet{conference:AFrameworkForTheStructuralAnalysisOfRestApis}, they conducted a structural analysis of 286 real-world Web APIs. 
In detail, the framework takes an interface description, converts it into a canonical metamodel, and calculates several metrics to support API governance.
As a usage example, they demonstrated how calculated metrics can be used to estimate the user-perceived complexity of an API, which is related to API understandability.
For this, they randomly selected 10 of their 286 APIs and let 9 software developers rank them based on their perceived complexity. 
Based on their knowledge and experience, the authors then defined several metrics for user-perceived complexity and calculated these metrics for the 10 APIs. 
As a result, some calculated metrics coincided with the developers' judgments and were proposed for automatic complexity estimation.
A follow-up confirmatory study with a larger sample size to substantiate this exploratory approach is missing so far.

An approach similar to~\citet{Haupt2018} was used in a study by ~\citet{Bogner2019}.
They proposed a modular framework, the \textit{RESTful API Metric Analyzer} (RAMA), which calculates maintainability metrics from interface descriptions and enables the automatic evaluation of Web APIs. 
More precisely, RAMA converts an interface description into a hierarchical model and calculates 10 service-based maintainability metrics.  
In a benchmark run, the authors applied RAMA to a set of 1,737 real-world APIs, and calculated quartile-based thresholds for the metrics.
However, relating the metrics to other software quality correlates is missing to evaluate their effectiveness.

The impact of good and poor Web API design on understandability and reusability has been investigated in a series of publications by Palma et al.: 
in~\citep{article:Palma2017}, the authors defined 12 linguistic patterns and antipatterns focusing on URI design in Web APIs, which may impact understandability and reusability of such APIs.
Moreover, they proposed algorithms for their detection and implemented them as part of the \textit{Service-Oriented Framework for Antipatterns} (SOFA).
They used SOFA to detect linguistic \mbox{(anti-)patterns} in 18 real-world APIs, with the result that most of the investigated APIs used appropriate resource names and did not use verbs within URI paths. 
However, URI paths often did not convey hierarchical structures. 

In another study, \citet{conference:Palma2021} tried to answer whether a well-designed RESTful API also has good linguistic quality and, vice versa, whether poorly designed Web APIs have poor linguistic quality. 
For this, they used SOFA to analyze 8 Google APIs and to detect 9 design patterns and antipatterns, as well as 12 linguistic patterns and antipatterns. 
However, their statistical tests revealed only negligible relationships between RESTful design and linguistic design qualities.  

Subsequently, they extended SOFA with further linguistic \mbox{(anti-)patterns}, improved approaches for their detection, and applied the linguistic quality analysis on Web APIs from the IoT domain~\citep{article:Palma2022IoT} or compared the quality between public, partner, and private APIs~\citep{conference:Palma2022PuPaPr}.

In summary, existing studies assessed the quality of real-world Web APIs by collecting metrics and detecting (anti-)patterns.
The latter are somewhat related to RESTful API design rules and best practices that should, in theory, improve several quality aspects of an API. 
However, in many cases, there is no empirical evidence for the effectiveness of the impact of these design rules and best practices on software quality, especially on understandability.
\citet{Haupt2018} and \citet{Kotstein2021} used subjective ratings, but no studies in which human participants solve comprehension or maintenance tasks have been conducted.
To the best of our knowledge, our experiment is the first study that investigated the understandability impact of violating RESTful API design rules from the perspective of human API consumers.

\section{Research Design}
In this section, we describe the details of our methodology.
We roughly follow the reporting structure for software engineering experiments proposed by \citet{Jedlitschka2008}.
Inspired by the experiment characteristics discussed by \citet{Wyrich2022}, Table~\ref{tab:experimentOverview} provides a quick overview of the most important characteristics of the study.
For transparency and reproducibility, we publish our experiment artifacts on Zenodo\footnote{\url{https://doi.org/10.5281/zenodo.7381500}}.

\subsection{Research Questions}
We investigated three different research questions in this study.

\smallskip

\noindent\textbf{RQ1:} Which design rules have a significant impact on the understandability of Web APIs?

\smallskip

Our hypothesis for this central, confirmatory RQ was that each selected rule should improve understandability, i.e., the effectiveness and efficiency of grasping the functionality and intended purpose of a Web API endpoint.

\smallskip

\noindent\textbf{RQ2:} Which design rules have a significant impact on software professionals' perceived difficulty while understanding Web APIs?

\smallskip

For the confirmatory part of this RQ, we hypothesized that API snippets with rule violations are rated as more difficult to understand.
Additionally, we analyzed the correlation between actual and perceived understandability in an exploratory part to identify potential differences.

\smallskip

\noindent\textbf{RQ3:} How do participant demographics influence the effectiveness and perception of design rules for understanding Web APIs?

\smallskip

We did not have strong hypotheses for this exploratory RQ.
Nonetheless, we had some intuitions about attributes that may be interesting to analyze.
For example, it could be possible that adhering to the design rules mostly has an influence on experienced professionals but not on students (or vice versa).
Furthermore, some rules might require the participant to know about the Richardson maturity model or some rule violations might be perceived as more critical by participants from academia or from industry.
During the study design phase, we selected some general demographic attributes plus several specific ones for the experiment context.

\begin{table}[H]
    \caption{Experiment overview}
    \label{tab:experimentOverview}
    \centering
    \begin{tabular}{l p{0.62\textwidth}}
        \hline
        \hline
        Goal & Study the impact of design rules on the understandability of Web APIs\\
        Study objects & 12 design rules compiled from \citet{Masse2011} and \citet{Richardson2007}, two functionally equivalent API snippets per rule (one follows the rule, one violates it)\\
        Participants & 105 people with at least basic REST-related experience (both students and professionals, both from academia and industry)\\
        Setting & Online experiment via LimeSurvey\\
        Tasks & Answering comprehension question about API snippets for RQ1 (12 per participant), rating the difficulty to understand an API snippet for RQ2 (12 per participant)\\
        Dependent variables & Timed actual understandability (TAU) for RQ1, perceived difficulty rating for RQ2\\
        Treatments & API snippet follows a \textit{rule} (version 1) or API snippet contains a \textit{violation} of a rule (version 2)\\
        Other independent variables & Demographic attributes like REST-related experience or current role (RQ3)\\
        Design & Hybrid between crossover and between-subjects design (2 counterbalanced sequences, each with 6 tasks for \textit{rule} and 6 for \textit{violation})\\
        \hline
        \hline
    \end{tabular}
\end{table}

\subsection{Participants and Sampling}
The only requirements for participation were basic knowledge of REST and HTTP, as well as the ability to understand English.
Our goal was to attract participants from diverse backgrounds and experience levels, e.g., both students and professionals, both participants from industry and academia, etc.
We used \textit{convenience sampling} mixed with \textit{referral-chain sampling}, i.e., we distributed the call for participation within our personal networks via email, and kindly asked for forwarding to relevant circles~\citep{Baltes2022}.
A similar message was displayed after the experiment to encourage sharing.
Students were recruited via internal mailing lists of several universities.
Moreover, we advertised the study via social media, such as Twitter\footnote{\url{https://www.twitter.com}}, LinkedIn\footnote{\url{https://www.linkedin.com}}, XING\footnote{\url{https://www.xing.com}}, and in several technology-related subreddits\footnote{\url{https://www.reddit.com}}.

\subsection{Experiment Objects}
In this experiment, the objects under study were 12 design rules for RESTful APIs that have been proposed in the literature.
They are summarized in Table~\ref{tab:rules} together with short identifiers that we use throughout the rest of the paper.
Rule selection was guided by the results of our previous Delphi study~\citep{Kotstein2021}, i.e., we focused on rules from \citet{Masse2011} that were perceived as very important by industry experts, with an influence on maintainability or usability.
Additionally, we included three instances of the \texttt{PathHierarchy} rule proposed by \citet{Richardson2007}.
While this rule was not part of our previous study, it has strong relationships to Massé's rules \textit{Variable path segments may be substituted with identity-based values} and \textit{Forward slash separator (/) must be used to indicate a hierarchical relationship}.
Both of these rules fulfill the above criteria, i.e., high importance plus influence on maintainability or usability.
In the following, the rule descriptions are taken from \citet{Masse2011} and \citet{Richardson2007} respectively.
For each rule, we also present the concrete endpoint pair that was used in the experiment, one version for following the rule and one for violating it.
These concrete Web API examples and the \textit{violation} versions are based on our industry experience, but also on existing public APIs that we identified via the APIs Guru repository\footnote{\url{https://apis.guru}}.
For our experiment, the chosen real-world examples were adapted to simplify them and to avoid that participants are already familiar with the presented HTTP endpoints.
During the pilot, we discussed the created pairs of \textit{rule} and \textit{violation} with external experts to validate if the \textit{violation} snippets were not strongly exaggerated.
Several snippets were adapted based on this feedback, and one task was dropped entirely.

\begin{table}[ht]
    \caption{Selected design rules for RESTful APIs}
	\label{tab:rules}
	\centering
	\begin{tabular}{
	    l
	    >{\raggedright\arraybackslash}p{0.37\textwidth}
	    >{\raggedright\arraybackslash}p{0.1708\textwidth}
	    >{\raggedright\arraybackslash}p{0.1903\textwidth}
	}
		ID & Rule & Source & Category\\
		\hline
		\hline
		\texttt{PluralNoun} & A plural noun should be used for collection and store names & \citet{Masse2011} & URI Design \\
        \texttt{VerbController} & A verb or verb phrase should be used for controller names & \citet{Masse2011} & URI Design \\
        \texttt{CRUDNames} & CRUD function names should not be used in URIs & \citet{Masse2011} & URI Design \\
        \makecell[l]{\texttt{PathHierarchy}\\(3 versions)} & Use path variables to encode hierarchy & \citet{Richardson2007} & Hierarchy Design \\
        \texttt{NoTunnel} & \texttt{GET} and \texttt{POST} must not be used to tunnel other request methods & \citet{Masse2011} & Request Methods \\
        \texttt{GETRetrieve} & \texttt{GET} must be used to retrieve a representation of a resource & \citet{Masse2011} & Request Methods \\
        \texttt{POSTCreate} & \texttt{POST} must be used to create a new resource in a collection & \citet{Masse2011} & Request Methods \\
        \texttt{NoRC200Error} & \texttt{200 (OK)} must not be used to communicate errors in the response body & \citet{Masse2011} & HTTP Status Codes \\
        \texttt{RC401} & \texttt{401 (Unauthorized)} must be used when there is a problem with the client’s credentials & \citet{Masse2011} & HTTP Status Codes \\
        \texttt{RC415} & \texttt{415 (Unsupported Media Type)} must be used when the media type of a request's payload cannot be processed & \citet{Masse2011} & HTTP Status Codes \\
		\hline
		\hline
	\end{tabular}
\end{table}

\subsubsection{URI Design}
The three rules in this category are concerned with the concrete design of URI paths in an API.

\paragraph{PluralNoun:}
\citet{Masse2011} defines this rule for both \textit{collections} and \textit{stores}, which we merge into a single rule for simplicity.
In both cases, it prescribes to use a plural noun as the name in the URI.
A violation of this rule would be to use a singular noun for a \textit{collection} or \textit{store} name instead.

\noindent
\begin{tabular}{ll}
Rule: & \texttt{GET /groups/\{groupId\}/members}\\
Violation: & \texttt{GET /groups/\{groupId\}/member}
\end{tabular}

\paragraph{VerbController:}
A controller provides an action that cannot be easily mapped to a typical CRUD operation on a resource.
In relation to function names in source code, \citet{Masse2011} proposes to always use a verb or verb phrase for controller resources.
Using a noun instead would be a violation.

\noindent
\begin{tabular}{ll}
Rule: & \texttt{POST /servers/\{serverId\}/backups/\{backupId\}/restore}\\
Violation: & \texttt{POST /servers/\{serverId\}/backups/\{backupId\}/restoration}
\end{tabular}

\paragraph{CRUDNames:}
Based on the invoked HTTP method, a RESTful API selects the semantically equivalent CRUD operation to perform.
Therefore, \citet{Masse2011} prescribes not to use CRUD function names like \enquote{create} or \enquote{update} in URIs, especially not with incorrect HTTP verbs.
Adhering to this rule means solely relying on the HTTP verb to indicate the wanted CRUD operation.
Our chosen example also includes the rule \enquote{\texttt{DELETE} must be used to remove a
resource from its parent}~\citep{Masse2011} for the violation.

\noindent
\begin{tabular}{ll}
Rule: & \texttt{DELETE /messaging-topics/\{topicId\}/queues/\{queueId\}}\\
Violation: & \texttt{GET /messaging-topics/\{topicId\}/delete-queue/\{queueId\}}
\end{tabular}

\subsubsection{Hierarchy Design}
A prominent rule from \citet{Richardson2007} prescribes the use of path parameters to encode the hierarchy of resources in a URI.
Since there are several possibilities to apply and interpret this, we created three specific rules based on this idea (\texttt{PathHierarchy1} to \texttt{PathHierarchy3}).

\paragraph{PathHierarchy1 (path params vs. query params):}
Version 1 explores the difference between using path parameters (\textit{rule}) and query parameters (\textit{violation}) for retrieving a hierarchically structured resource.

\noindent
\begin{tabular}{ll}
Rule: & \texttt{GET /shops/\{shopId\}/products/\{productId\}}\\
Violation: & \texttt{GET /shops/products?shopId=\{shopId\}\&productId=\{productId\}}
\end{tabular}

\paragraph{PathHierarchy2 (top-down vs. bottom-up):} In version 2, the difference between structuring the hierarchy top-down / from left to right (\textit{rule}) and bottom-up / from right to left (\textit{violation}) is tested.

\noindent
\begin{tabular}{ll}
Rule: & \texttt{GET /companies/\{companyId\}/employees}\\
Violation: & \texttt{GET /employees/companies/\{companyId\}}
\end{tabular}

\paragraph{PathHierarchy3 (hierarchical path vs. short path):} Lastly, version 3 analyzes differences during the creation of a resource when either using a long hierarchical path with parameters (\textit{rule}) or a short path with parameters in the request body (\textit{violation}).

\noindent
\begin{tabular}{ll}
Rule: & \texttt{POST /customers/\{customerId\}/environments/\{environmentId\}/servers}\\
Violation: & \texttt{POST /servers}
\end{tabular}

\subsubsection{Request Methods}
In accordance with level 2 of the Richardson maturity model, \citet{Masse2011} proposes several rules that prescribe that each HTTP method should exclusively be used for its semantically equivalent operation.
We selected three of these rules for our experiment.

\paragraph{NoTunnel:}
One of these rules states that \texttt{GET} and \texttt{POST} must not be used to tunnel other request methods, which might seem tempting for the sake of simplicity.
In our example, an API correctly uses \texttt{PUT} to update a resource (\textit{rule}), whereas the \textit{violation} always uses \texttt{POST} and tunnels the update operation via an additional query parameter.

\noindent
\begin{tabular}{ll}
Rule: & \texttt{PUT /trainings/\{trainingId\}/organizers/\{organizerId\}}\\
Violation: & \texttt{POST /trainings/\{trainingId\}/organizers/\{organizerId\}?operation=update}
\end{tabular}

\paragraph{GETRetrieve:}
In similar fashion, another rule from \citet{Masse2011} states that the HTTP method \texttt{GET} must be used to retrieve a representation of a resource.
Since \texttt{GET} requests have no request body, it may seem tempting to use \texttt{POST} in some cases to be able to use a JSON object instead of overly complex query parameters.
A typical example of this is a search resource.
Adhering to the \textit{rule} requires using \texttt{GET} plus query parameters, while reverting to \texttt{POST} plus a request body with the search options is a \textit{violation}.

\noindent
\begin{tabular}{ll}
Rule: & \texttt{GET /events?date=2022-10-03\&category=music}\\
Violation: & \texttt{POST /events/search}
\end{tabular}

\paragraph{POSTCreate:}
Lastly, we tested the complementary rule for \texttt{POST}, namely that this method must be used to create resources in a collection~\citep{Masse2011}.
A typical violation of this rule is the use of \texttt{PUT} to create a resource, as seen in our chosen example.

\noindent
\begin{tabular}{ll}
Rule: & \texttt{POST /customers/\{customerId\}/orders}\\
Violation: & \texttt{PUT /customers/\{customerId\}/orders}
\end{tabular}

\subsubsection{HTTP Status Codes}

Another important theme for RESTful API design is the correct usage of HTTP status codes with response messages.
\citet{Masse2011} provides a number of rules in this area, from which we selected three in total, namely the rules for \texttt{200 (OK)}, \texttt{401 (Unauthorized)}, and \texttt{415 (Unsupported Media Type)}.
Contrary to the previous three categories, the two versions of our used examples (\textit{rule} and \textit{violation}) do not differ in the displayed endpoint, but only for the displayed response including the response code.

\paragraph{NoRC200Error:}
The first rule in this category states that the response code \texttt{200 (OK)} must not be used in case of error.
Instead, client-side (\texttt{4XX}) or server-side (\texttt{5XX}) error codes must be used accordingly, based on the nature of the error.

\noindent
\begin{tabular}{lp{9.5cm}}
Rule: & A required parameter is missing in the request body. The server indicates there was a problem and correctly responds with \texttt{400 (Bad Request)}.\\
Violation: & A required parameter is missing in the request body. The server indicates there was a problem, but incorrectly responds with \texttt{200 (OK)}.\\
\end{tabular}

\paragraph{RC401:}
A similar rule prescribes the use of the status code \texttt{401 (Unauthorized)} in case of issues with client credentials, e.g., during a login attempt.
Using a different client-side response code like \texttt{400 (Bad Request)} or \texttt{403 (Forbidden)} should be avoided for such errors.
Since the nuances between \texttt{401} (not logged in or login failed) and \texttt{403} (logged-in user does not have the required privileges) might not be fully clear to all participants, we opted for the more generic code \texttt{400 (Bad Request)} in the violation example.

\noindent
\begin{tabular}{lp{9.5cm}}
Rule: & A secured resource is accessed with an empty \texttt{Bearer} token in the \texttt{Authorization} header. The server indicates there was a problem and correctly responds with \texttt{401 (Unauthorized)}.\\
Violation: & A secured resource is accessed with an empty \texttt{Bearer} token in the \texttt{Authorization} header. The server indicates there was a problem, but incorrectly responds with \texttt{400 (Bad Request)}.\\
\end{tabular}

\paragraph{RC415:}
The final rule in this category focuses on the correct usage of the status code \texttt{415 (Unsupported Media Type)}, which must be returned if the client uses a media type for the request body that cannot be processed by the server.
A typical example is a request body in XML when the server only supports JSON.
Returning a different client-side error code, e.g., \texttt{400 (Bad Request)}, should be avoided in this case.

\noindent
\begin{tabular}{lp{9.5cm}}
Rule: & A request body contains XML, even though the server only accepts JSON. The server indicates there was a problem and correctly responds with \texttt{415 (Unsupported Media Type)}.\\
Violation: & A request body contains XML, even though the server only accepts JSON. The server indicates there was a problem, but incorrectly responds with \texttt{400 (Bad Request)}.\\
\end{tabular}

\subsection{Material}
\begin{figure}[H]
  \centering
  \begin{minipage}[b]{0.496\textwidth}
	\includegraphics[width=\textwidth]{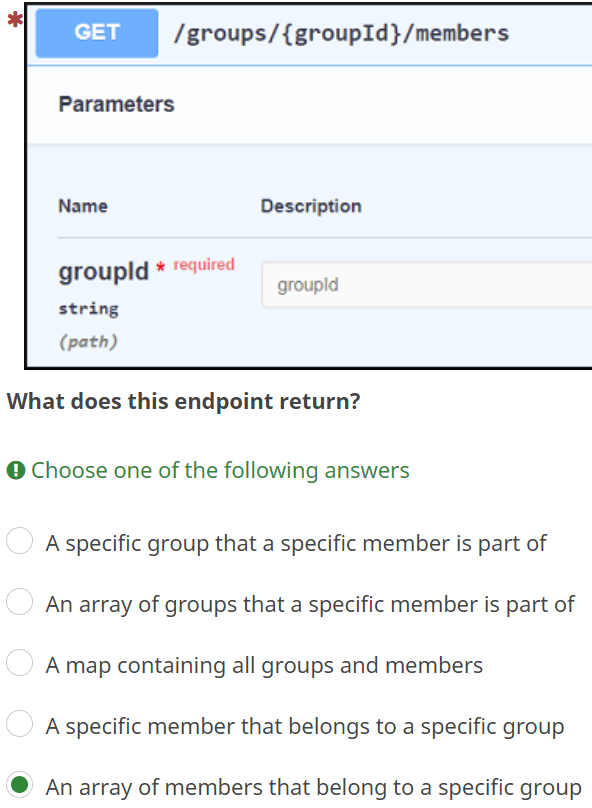}
  \end{minipage}
  \hfill
  \begin{minipage}[b]{0.496\textwidth}
    \includegraphics[width=\textwidth]{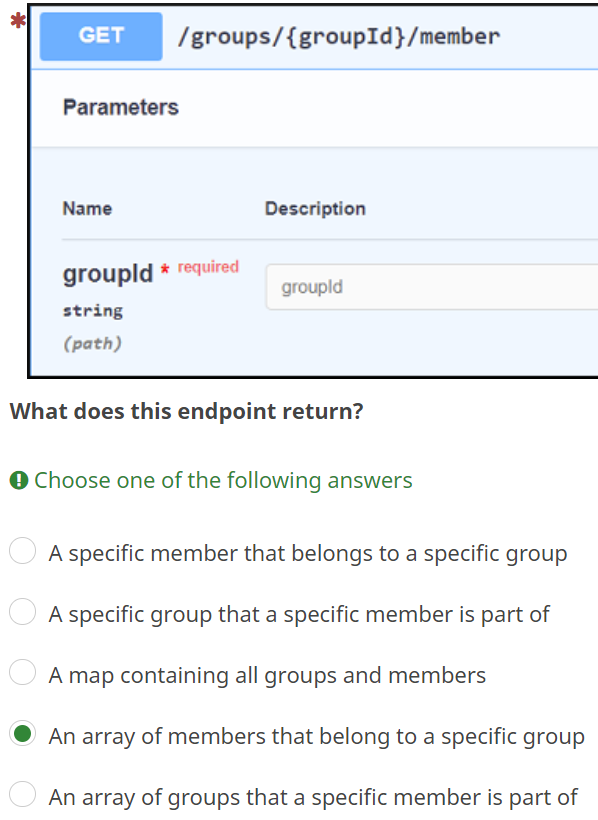}
  \end{minipage}
  \caption{Example of a return value question based on the rule \texttt{PluralNoun} (\textit{rule} on the left, \textit{violation} on the right, correct answer is checked)}
  \label{fig:questionTypeReturn}
\end{figure}

To incorporate the created API rule examples into our experiment, we relied on the OpenAPI specification format\footnote{\url{https://www.openapis.org}}, one of the most popular ways to document Web APIs many practitioners and researchers should be familiar with.
Using the Swagger editor\footnote{\url{https://editor.swagger.io}}, we created two OpenAPI documents, one with the examples following the rules and one with the violation examples.
Each document contained 12 endpoints, one per rule.
We then created screenshots of the graphical representation of each resource, with the purpose of showing them to participants with each task (see, e.g., Fig.~\ref{fig:questionTypeReturn}).

To reach a larger and more diverse audience, we decided to conduct an online experiment via a web-based tool.
We selected the open-source survey tool LimeSurvey\footnote{\url{https://www.limesurvey.org}} for this purpose, as it provides all the features we need.
Additionally, we had access to an existing LimeSurvey instance via one of our universities.
It supports several types of questions, is highly customizable, and also allows measuring the duration per task, i.e., survey question, which we needed for our experiment.
Lastly, random assignment of participants to sequences is also possible.
This setup meant that participants exclusively used LimeSurvey for the experiment via a computing device and web browser of their choice.
All necessary information was provided this way.

\begin{figure}[H]
  \centering
  \begin{minipage}[b]{0.485\textwidth}
	\includegraphics[width=\textwidth]{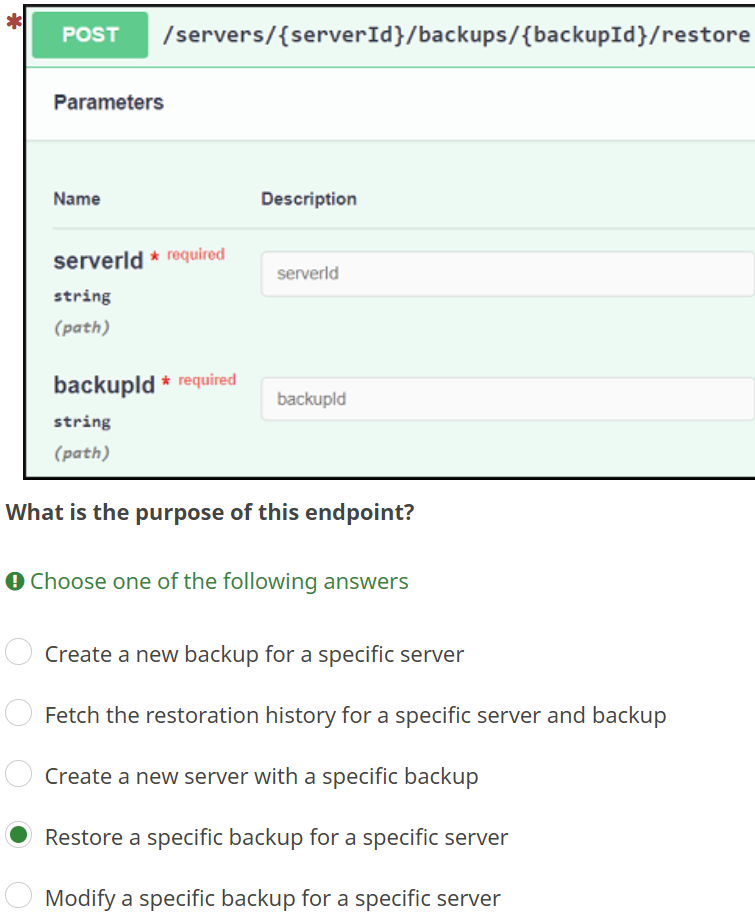}
  \end{minipage}
  \hfill
  \begin{minipage}[b]{0.505\textwidth}
    \includegraphics[width=\textwidth]{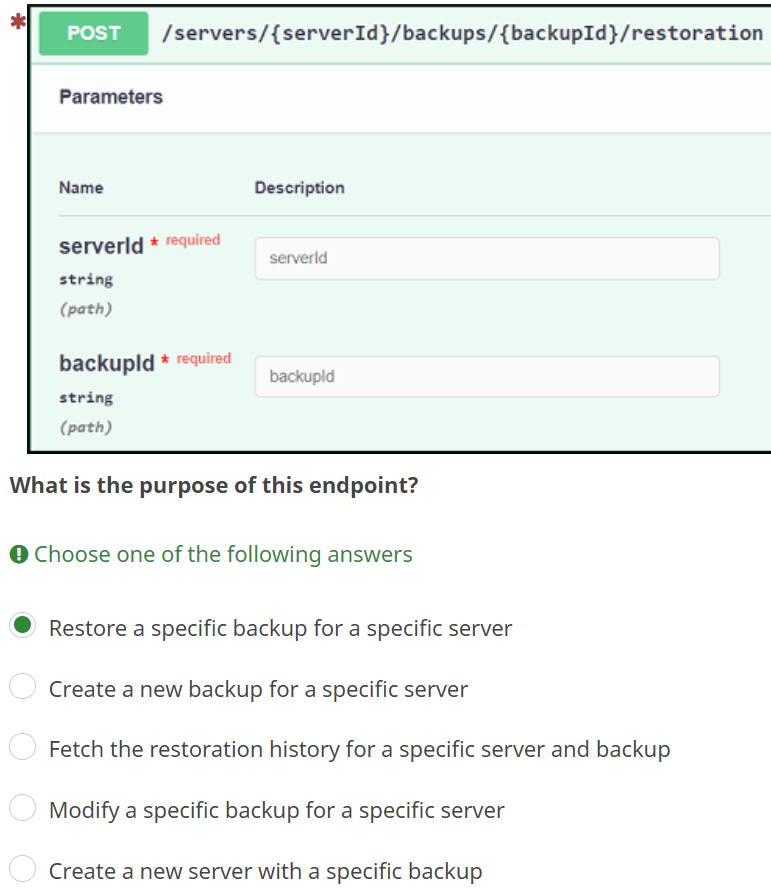}
  \end{minipage}
  \caption{Example of an endpoint purpose question based on the rule \texttt{VerbController} (\textit{rule} on the left, \textit{violation} on the right, correct answer is checked)}
  \label{fig:questionTypePurpose}
\end{figure}

\begin{figure}[H]
  \centering
  \begin{minipage}[b]{0.476\textwidth}
	\includegraphics[width=\textwidth]{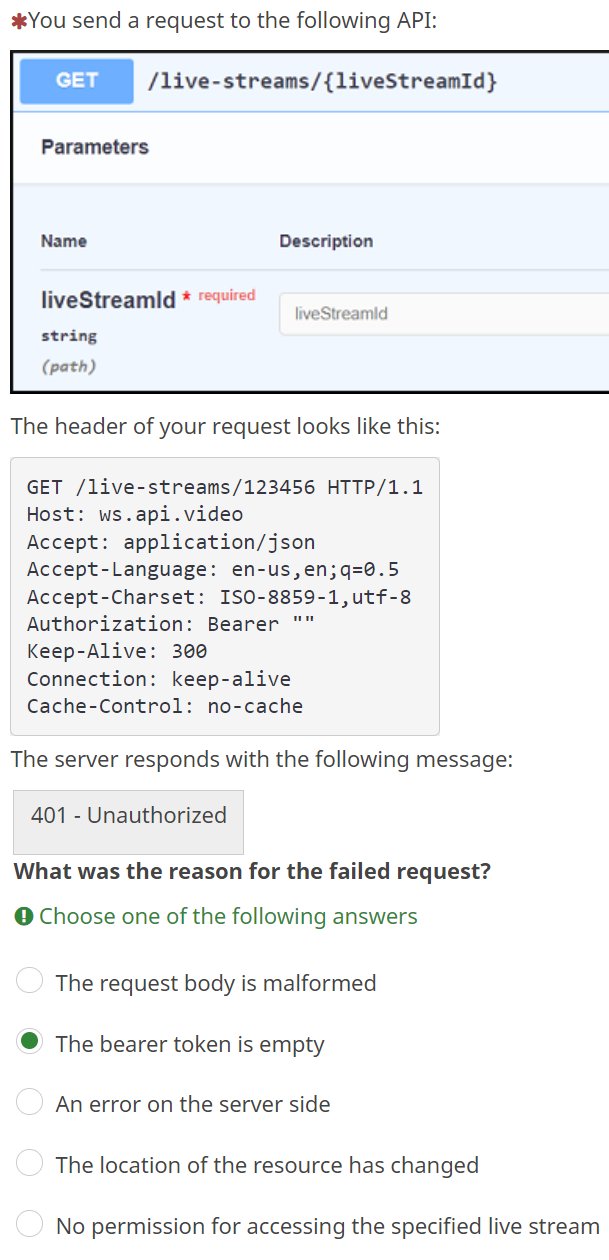}
  \end{minipage}
  \hfill
  \begin{minipage}[b]{0.514\textwidth}
    \includegraphics[width=\textwidth]{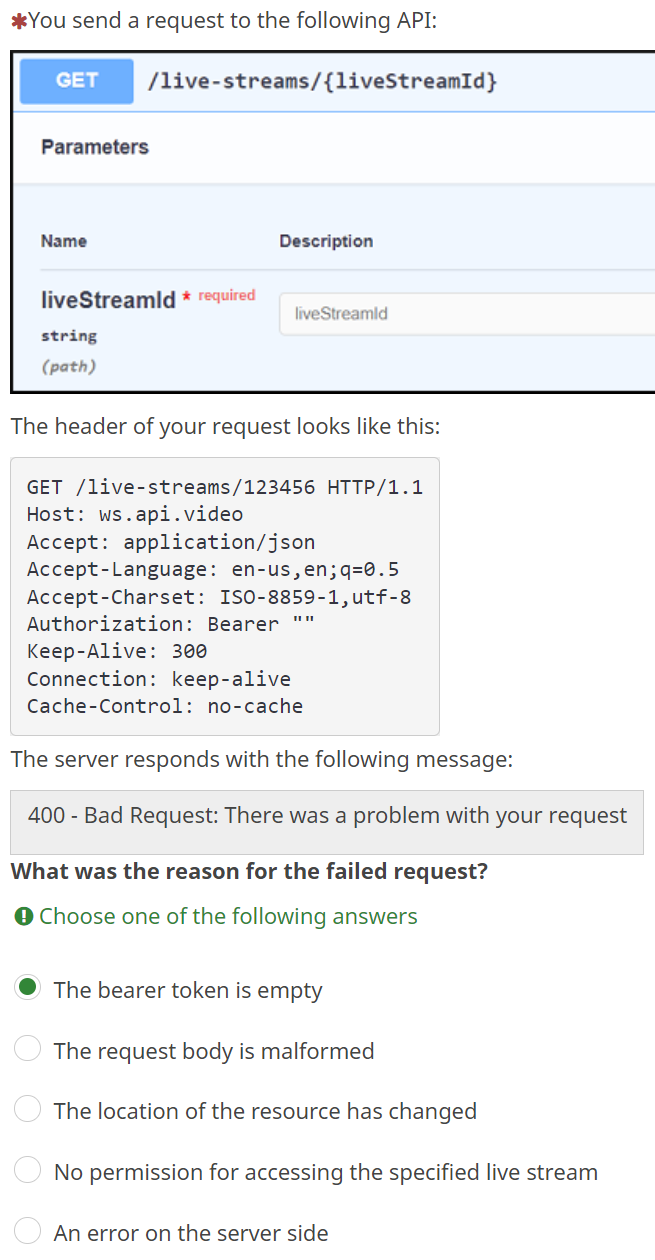}
  \end{minipage}
  \caption{Example of a response code reason question based on the rule \texttt{RC401} (\textit{rule} on the left, \textit{violation} on the right, correct answer is checked)}
  \label{fig:questionTypeResponseCode}
\end{figure}

\subsection{Tasks}
The participants' main task was to inspect and understand several Web API snippets that were presented in the graphical representation of the Swagger editor.
To evaluate understanding, participants had to answer one comprehension question per snippet.
Each of these questions was a single choice question with five different options presented in a random order, with exactly one of them being correct.
The options were the same regardless of whether the version adhering to the rule or the one violating the rule was displayed.
Opting for single choice comprehension questions had several advantages.
While free-text answers might reflect a more in-depth understanding of participants, they are much harder to correct.
Additionally, writing free-text answers takes more time and effort for participants, which may increase the drop-out rate.
Lastly, participants may specify answers with different level of details.
This not only complicates the grading, but it also influences the time to answer, which we include in our comprehension measures.
Depending on the shown snippet, exactly one of three different types of comprehension questions was asked.

\noindent\textbf{Return Value:}
Participants had to determine the return value of an endpoint, i.e., the type of entity and if a single object or a collection was returned.
This question type was used for snippets with \texttt{GET} requests, namely \texttt{PluralNoun}, \texttt{PathHierarchy1}, and \texttt{PathHierarchy2}.
An example is shown in Fig.~\ref{fig:questionTypeReturn}.

\noindent\textbf{Endpoint Purpose:}
Participants had to determine the purpose of the shown endpoint, i.e., what operation or functionality was executed on invocation.
This question type was used for snippets which do not always return entities, e.g., \texttt{POST}, \texttt{PUT}, and \texttt{DELETE} requests, namely \texttt{VerbController}, \texttt{CRUDNames}, \texttt{PathHierarchy3}, \texttt{NoTunnel}, \texttt{GETRetrieve}, and \texttt{POSTCreate}.
An example is presented in Fig.~\ref{fig:questionTypePurpose}.

\noindent\textbf{Response Code Reason:}
Participants had to determine why a certain request failed or what the outcome of a request was.
This question type was exclusively used for snippets of the category \textit{HTTP Status Codes}, namely \texttt{NoRC200Error}, \texttt{RC401}, and \texttt{RC415}.
An example is presented in Fig.~\ref{fig:questionTypeResponseCode}.

After each comprehension question, participants also had to rate the perceived difficulty of understanding the API snippet on a 5-point ordinal scale, ranging from \textit{very easy} (1) to \textit{very hard} (5).
For this purpose, the last snippet was shown again, so participants did not have to rely solely on memory.

\subsection{Variables and Hypotheses}
\label{sec:varAndHypo}
Two \textbf{dependent variables} were used in this experiment.
For RQ1, we needed a metric for understandability.
To operationalize this quality attribute, we collected both the \textit{correctness} and \textit{duration} for each task per participant.
Since every task had a single correct answer, \textit{correctness} was a binary variable, with 0 for false and 1 for correct.
The required \textit{duration} per task was documented in seconds.
To combine these two measures into a single variable, we adapted an aggregation procedure from \cite{Scalabrino2021}, namely \textit{Timed Actual Understandability} (TAU).
In our experiment, TAU for a participant $p$ and task $t$ was calculated as follows:
\begin{equation}
TAU\textsubscript{p,t} = correctness\textsubscript{p,t}\times (1-\frac{duration\textsubscript{p,t}}{max(durations\textsubscript{t})})
\end{equation}
TAU produces values between 0 and 1, with values closer to 1 indicating a higher degree of understandability.
For an incorrect answer, TAU is always 0.
For a correct answer, the task duration is set in relation to the maximum duration that was recorded for this task.
This is then inverted by subtracting it from 1, meaning the faster the correct answer was found, the greater is TAU.
As such, TAU represents a pragmatic aggregation of \textit{correctness} and \textit{duration} that respects differences between participants in the sample and leads to easily interpretable values.
Even though it leads to unusual distributions (see, e.g., Fig.~\ref{fig:stripplotTAU1}), we therefore chose TAU as the dependent variable for RQ1.
For RQ2, the dependent variable was the \textit{perceived difficulty}, i.e., the rating that participants had to give after each task using a 5-point ordinal scale, ranging from \textit{very easy} (1) to \textit{very hard} (5).

As the directly controlled \textbf{independent variable}, we used the version number of the respective snippet.
This was either version 1 (\textit{rule}) that followed the rule or version 2 (\textit{violation}) that violated the rule.
Additionally, we had uncontrollable independent variables that we collected for RQ3, namely various demographic attributes of participants such as their current role, years of experience with REST, or knowledge of the Richardson maturity model.

Based on these variables, we formulated hypotheses for the confirmatory questions RQ1 (difference in actual understandability between \textit{rule} and \textit{violation}) and RQ2 (difference in perceived understandability).
They are displayed in Table~\ref{tab:hypotheses}.
In both cases, we expected that the version following the design rule would lead to significantly better results in comparison to the version violating the rule.
While we only list two hypotheses, each of the 12 rules was tested individually to clearly identify which rules have an impact and which do not.
Finally, the exploratory RQ3 did not have a clear hypothesis.

\begin{table}[H]
    \caption{Null hypotheses with their alternatives for the two confirmatory RQs}
    \label{tab:hypotheses}
    \centering
    \small
    \begin{tabular}{l p{1.765cm} p{4cm} p{4cm}}
        RQ & Metric & Null Hypothesis & Alternative Hypothesis\\
        \hline
        \hline
        RQ1
        &
        Timed actual understandability (TAU)
        &
        \textbf{H$_{0}^{1}$:} Web API snippets adhering to design rules are equally or less understandable than snippets violating rules.
        & 
        \textbf{H$_{1}^{1}$:} Web API snippets adhering to design rules are more understandable than snippets violating rules.\\
        \hline
        RQ2
        &
        Perceived difficulty
        &
        \textbf{H$_{0}^{2}$:} Web API snippets adhering to design rules are rated as equally or more difficult to understand than snippets violating rules.
        & 
        \textbf{H$_{1}^{2}$:} Web API snippets adhering to design rules are rated as less difficult to understand than snippets violating rules.\\
        \hline
        \hline
    \end{tabular}
\end{table}

\subsection{Experiment Design}
We opted for a mixed experiment design that can be described as a hybrid between a \textit{crossover design}~\citep{Vegas2016} and a \textit{between-subjects design}~\citep{Wohlin2012}, thereby exploiting specific benefits of both.
In a crossover design (a special form of a within-subjects design), each participant receives each treatment at least once, but the order in which participants receive the treatments differs.
Our participants worked on six tasks with the \textit{rule} version and six tasks with \textit{violation}.
However, in typical crossover experiments in software engineering, each participant applies all treatments to a given experiment object~\citep{Wohlin2012}, which was not the case for our experiment.
Only half of our participants worked on each version of a specific task, which is the between-subject characteristic of our experiment.
This mixed design is still fairly robust regarding inter-participant differences because participants are not assigned to a single treatment.
Letting each participant work on both treatments for the same rule would be even more robust, but would also create problems with familiarization effects.
It also would increase the experiment duration considerably, leading to fewer participants and higher mortality or fewer rules that could be tested.
Lastly, this design still provides us with many observations to overall compare \texttt{rule} vs. \texttt{violation}.

To avoid suboptimal task orderings regarding treatments or task categories that could cause carryover or order effects, we did not randomize the order of tasks.
Instead, we consciously designed two sequences with counterbalanced task orders (see Table~\ref{tab:taskOrder}), which is common for crossover designs~\citep{Vegas2016}.
Our goals were to spread out the tasks of the same category, to ensure the same number of \texttt{rule} and \texttt{violation} tasks, to avoid having the same treatment too often in a row, and to ensure that \texttt{rule} and \texttt{violation} for the same snippet appear at the same position per sequence.
Participants were randomly assigned to one of these sequences at the start of the experiment.

\begin{table}[H]
    \caption{Task orders for the two experiment sequences}
    \label{tab:taskOrder}
    \centering
    \begin{tabular}{r l l l l}
        \# & Rule & Category & Sequence 1 & Sequence 2\\
        \hline
        \hline
        1 & \texttt{VerbController} & URI Design & violation & rule \\
        2 & \texttt{RC415} & HTTP Status Codes & violation & rule \\
        3 & \texttt{PathHierarchy2} & Hierarchy Design & rule & violation \\
        4 & \texttt{GETRetrieve} & Request Methods & violation & rule \\
        5 & \texttt{PathHierarchy3} & Hierarchy Design & rule & violation \\
        6 & \texttt{NoRC200Error} & HTTP Status Codes & violation & rule \\
        7 & \texttt{NoTunnel} & Request Methods & rule & violation \\
        8 & \texttt{PluralNoun} & URI Design & rule & violation \\
        9 & \texttt{POSTCreate} & Request Methods & violation & rule \\
        10 & \texttt{PathHierarchy1} & Hierarchy Design & violation & rule \\
        11 & \texttt{CRUDNames} & URI Design & rule & violation \\
        12 & \texttt{RC401} & HTTP Status Codes & rule & violation \\
        \hline
        \hline
    \end{tabular}
\end{table}

This experiment design was the result of several iterations with internal discussion, followed by a pilot with external reviewers.
Based on the pilot results, we refined the tasks and survey text.
Furthermore, we removed a few previously planned tasks to keep the required participation time below 15 minutes, leading to the total of 12 tasks.

\subsection{Experiment Execution}
Participation in the online experiment via LimeSurvey was open for a period of approximately three weeks.
At the start of and during this period, we actively promoted the experiment URL within in our network.
Participants starting the experiment were first presented with a welcome page containing some general information about the study.
We explained that our goal with this survey was to investigate the understandability of different Web API designs, and that basic REST knowledge was the only requirement for participation.
We mentioned that the experiment should take between 10-15 minutes and should ideally be finished in one sitting.
For each of the first 100 participants, we pledged to donate 1~\EUR{} to UNICEF\footnote{\url{https://www.unicef.org}} as additional motivation.

\begin{figure}[H]
    \centering
    \includegraphics[width=0.75\textwidth]{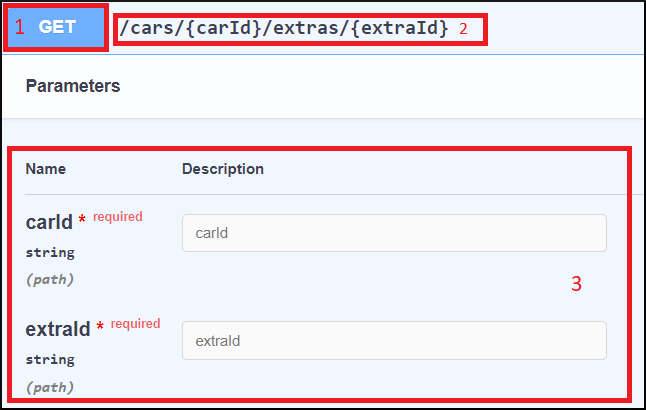}
    \caption{Example API snippet from the introduction page, explanations were provided for the numbered elements}
    \label{fig:introExample}
\end{figure}

Afterward, we described the tasks, namely that we will present 12 different API snippets.
For each snippet, they have to answer a multiple-choice question with five different options, e.g., \enquote{What is the purpose of this endpoint?}.
We would measure the time it takes to answer, but answering correctly would be more important.
After each comprehension question, they would rate how difficult it was to understand the API snippet without time measurements.
Finally, there would be a few demographic questions.
To familiarize participants with the used API snippet visualization, we also presented an example snippet with additional explanations about the different elements (see Fig.~\ref{fig:introExample}).

The last part of the welcome page contained our privacy policy.
We explained what data would be requested and that data would be used for statistical analysis and possibly be included in scientific publications.
Additionally, we emphasized that participation was strictly voluntary and anonymous, and that participants could choose to abandon the experiment at any time, resulting in the deletion of the previously entered data.
Participants had to consent to these terms before being able to start the experiment.

Upon accepting our privacy policy, participants started working on the tasks of their randomly assigned sequence (see Table~\ref{tab:taskOrder}).
For each of the 12 tasks, participants first analyzed the presented API snippet and answered the comprehension question.
Afterward, participants had to rate the difficulty of understanding the studied API snippet.
The snippet was displayed again for this purpose.
For the experiment part, participants answered a total of 24 questions, one comprehension and one rating question for each task.

Finally, participants answered the demographic questions, namely their country of origin, current role, technical API perspective (either API user / client developer, API developer / designer, or both), years of professional experience with REST, knowledge of the Richardson maturity model, and (if they knew about the model) their opinion about the minimal required maturity level a Web API should possess to be considered RESTful.
We also provided an optional free-text field for any final remarks or feedback participants wanted to give.
Finally, an outro page was presented, where we thanked participants and kindly asked them to forward the experiment URL to suitable colleagues.

\subsection{Experiment Analysis}
To analyze the experiment results, we first exported all responses as a CSV file.
We then performed the following data cleaning and transformation steps:

\begin{itemize}
    \item Removing 48 incomplete responses, i.e., participants that aborted the experiment at some point (rationale: avoid including responses without demographics, participants not fully committed, or participants who realized themselves that their background was not suitable)
    \item Resolving and harmonizing free-text answers for current role and technical API perspective (\enquote{Other:})
    \item Harmonizing country names
    \item Adding binary variables (\texttt{1} or \texttt{0}) for \texttt{is\_Student}, \texttt{is\_Academia} (an academic professional), and \texttt{is\_Germany}; if both \texttt{is\_Student} and \texttt{is\_Academia} were \texttt{0}, the participant was an industry practitioner
\end{itemize}

No substantial ambiguity was identified during the harmonization, and no final participant comments had to be considered for adapting the assigned correctness score.
However, since TAU is sensitive to outliers of the measured duration, we also analyzed the durations for all comprehension questions.
If answering the question took less than 5 seconds or more than 3 minutes, we judged this as invalid and removed the individual answer for all three dependent variables (correctness, time, and subjective rating).
The rationale for this was that we did not want to consider any responses (not even the subjective rating) where it was very likely that the participant had not concentrated fully on the task, e.g., by answering before reading everything or leaving the experiment for more than a few seconds to do something else.
Since questions of the type \textit{HTTP Status Codes} were more verbose, the allocated thresholds for these questions were 10 seconds and 4 minutes.
If these duration thresholds were triggered three or more times for the same participant, we removed their complete response instead of only the individual answers.
Due to this filtering, 2 complete responses (initially, we had 107 complete responses) and 12 individual answers were removed.
The number of valid responses considered in the analysis is displayed for each individual task and treatment in Table~\ref{tab:validResponses}.

\begin{table}[ht]
    \caption{Number of valid responses per individual task and treatment}
    \label{tab:validResponses}
    \centering
    \begin{tabular}{r l r r}
        \# & Rule & \# of responses for \textit{rule} & \# of responses for \textit{violation}\\
        \hline
        \hline
        1 & \texttt{VerbController} & 53 & 52\\
        2 & \texttt{RC415} & 53 & 52\\
        3 & \texttt{PathHierarchy2} & 52 & 53 \\
        4 & \texttt{GETRetrieve} & 53 & 52\\
        5 & \texttt{PathHierarchy3} & 52 & 52 \\
        6 & \texttt{NoRC200Error} & 53 & 50\\
        7 & \texttt{NoTunnel} & 50 & 52 \\
        8 & \texttt{PluralNoun} & 52 & 52 \\
        9 & \texttt{POSTCreate} & 53 & 51\\
        10 & \texttt{PathHierarchy1} & 53 & 50\\
        11 & \texttt{CRUDNames} & 51 & 53 \\
        12 & \texttt{RC401} & 51 & 53 \\
        \hline
        \hline
    \end{tabular}
\end{table}

The cleaned CSV file was then imported by an analysis script written in R\footnote{\url{https://www.r-project.org}}, which had also been tested and refined during the pilot study.
The script performs basic data transformation, calculates TAU, and provides general descriptive statistics as well as diagrams to visualize the data, such as box plots.
To select a suitable hypothesis test for RQ1 and RQ2, we first analyzed the data distributions with the Shapiro-Wilk test~\citep{Shapiro1965}.
For both dependent variables, the test resulted in a p-value $\ll 0.05$, i.e., the data did not follow a normal distribution, which called for a test without the assumption of normality.
We therefore opted for the non-parametric Wilcoxon-Mann-Whitney test~\citep{Neuhauser2011}, which has a mature R implementation in the \texttt{stats} package\footnote{\url{https://www.rdocumentation.org/packages/stats/versions/3.6.2/topics/wilcox.test}}.
To combat the multiple comparison problem (in our case, the testing of 12 rules), we applied the Holm-Bonferroni correction~\citep{Shaffer1995}.
We used the \texttt{p.adjust()} function from the \texttt{stats} package\footnote{\url{https://www.rdocumentation.org/packages/stats/versions/3.6.2/topics/p.adjust}} to adjust the computed p-values.
When an adjusted p-value was less than our targeted significance level of $0.05$, we rejected the null hypothesis and accepted the alternative.
To judge the effect size of accepted hypotheses, we additionally calculated Cohen's $d$~\citep{Cohen1988} using the \texttt{effectsize} package\footnote{\url{https://www.rdocumentation.org/packages/effectsize/versions/0.8.2}}.
Following \citet{Sawilowsky2009}, the values can be interpreted as follows:
\begin{itemize}
    \item $d < 0.2$: very small effect
    \item $0.2 \le d < 0.5$: small effect
    \item $0.5 \le d < 0.8$: medium effect
    \item $0.8 \le d < 1.2$: large effect
    \item $1.2 \le d < 2.0$: very large effect
    \item $d \ge 2.0$: huge effect
\end{itemize}

For the exploratory RQ3, we first used a correlation matrix for visual exploration.
Correlations of identified variable pairs were then further analyzed with Kendall's Tau~\citep{Kendall1938}, as it is more robust and permissive regarding assumptions about the data than other methods.
Lastly, we used linear regression\footnote{\url{https://www.rdocumentation.org/packages/stats/versions/3.6.2/topics/lm}} to further analyze combined effects of demographic attributes and to explore the potential for predictive modelling.

\section{Results and Discussion}
\label{sec:results}
In this section, we first present some general statistics about our participants.
Afterward, we provide the results for each RQ, starting with descriptive statistics and then presenting the hypothesis testing or correlation results.

\vspace{-0.1cm}
\subsection{Participant Demographics}
After the cleaning procedure, we were left with 105 valid responses, which is more than most software engineering experiments have.\footnote{See, e.g., the mapping study on code comprehension experiments by \cite{Wyrich2022}: median number of participants was 34 (for journals alone, it was 61)}
96 of these were complete, while we had to exclude one or two answers for 9 of them.
52 participants were randomly assigned to sequence 1 and 53 to sequence 2.
Table~\ref{tab:participants} compares the attributes for the two sequences, which are pretty similar in most cases.
Overall, our 105 participants had between 1 and 15 years of experience with REST, with a median of 4 years.
59 participants were from industry (56\%), 18 were professionals from academia (17\%), and 28 were students (27\%).

\begin{table}[H]
    \caption{Participant demographics by sequence (RMM: Richardson Maturity Model)}
    \label{tab:participants}
    \centering
    \begin{tabular}{
        r
        r
        >{\raggedleft\arraybackslash}p{0.16265\textwidth}
        >{\raggedleft\arraybackslash}p{0.1477\textwidth}
        >{\raggedleft\arraybackslash}p{0.157\textwidth}
        >{\raggedleft\arraybackslash}p{0.0982\textwidth}
        >{\raggedleft\arraybackslash}p{0.1492\textwidth}
        }
        Sequence & \# & REST exp. in years (median) & \# of industry participants & \# of academic participants & \# of students & \# with RMM knowledge \\
        \hline
        \hline
        1 & 52 & 3.5 & 31 (60\%) & 9 (17\%) & 12 (23\%) & 9 (17\%)\\
        2 & 53 & 4 & 28 (53\%) & 9 (17\%) & 16 (30\%) & 19 (36\%)\\
        \hline
        Total & 105 & 4 & 59 (56\%) & 18 (17\%) & 28 (27\%) & 28 (27\%)\\
        \hline
        \hline
    \end{tabular}
\end{table}

Most of our participants, namely 70 of 105 (67\%), were located in Germany, followed by Portugal (12), the US (6), and Switzerland (5).
The remaining 12 responses were distributed across 7 countries with between 1 and 3 participants.
Regarding the roles of the 59 industry participants, the majority of them were software engineers (41).
Eight were consultants, two were software architects, and two were team managers.
The remaining six each had a role that was only mentioned once, e.g., test engineer.
Concerning the technical API perspective, most participants reported to be both active in API development and usage (76), while 16 exclusively were API users / client developers, 8 exclusively were API developers, and 5 did not provide an answer to this optional question.
Interestingly, only 28 participants (27\%) reported knowing the Richardson maturity model.
When these 28 were then asked their opinion about the minimal maturity level that a Web API should have to be RESTful, 22 chose level 2 (79\%) and 6 level 3 (21\%).
No one selected level 0 and 1.
This may be an indication that the Richardson maturity model is not very well known and that HATEOAS is not perceived as an important requirement for RESTfulness by most professionals.
Especially the latter is in line with previous findings~\citep{Kotstein2021}.

\vspace{-0.2cm}
\subsection{Impact on Understandability (RQ1)}
\begin{figure}[H]
    \centering
    \includegraphics[width=\textwidth, trim={0.2cm 0.3cm 1.4cm 0.4cm}, clip]{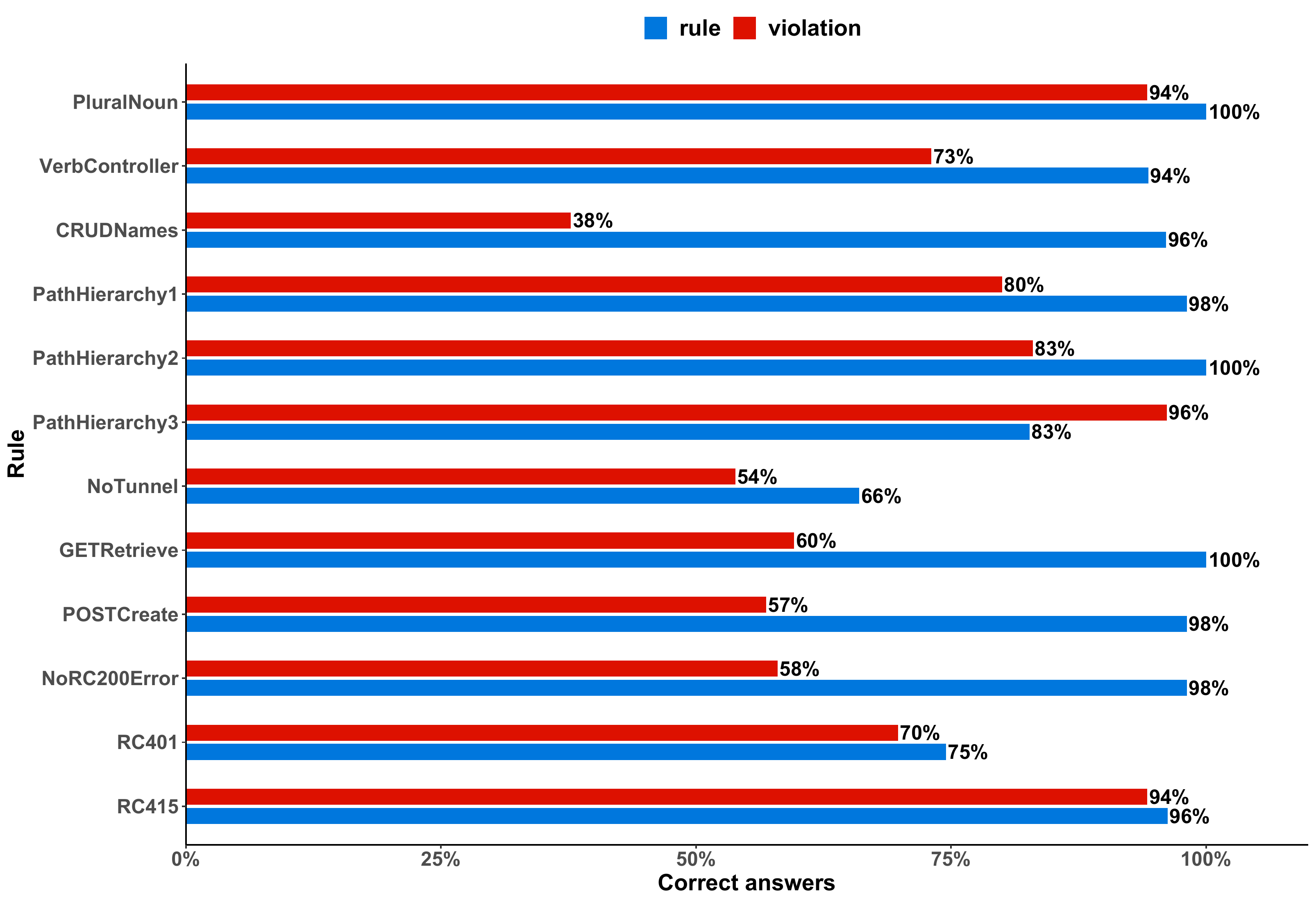}
    \vspace{-0.2cm}
    \caption{Comparison of correctness between treatments \textit{rule} and \textit{violation} (higher is better)}
    \label{fig:barplotPercent}
\end{figure}

\begin{figure}[H]
    \centering
    \includegraphics[width=\textwidth, trim={0.2cm 0.3cm 1.4cm 0.4cm}, clip]{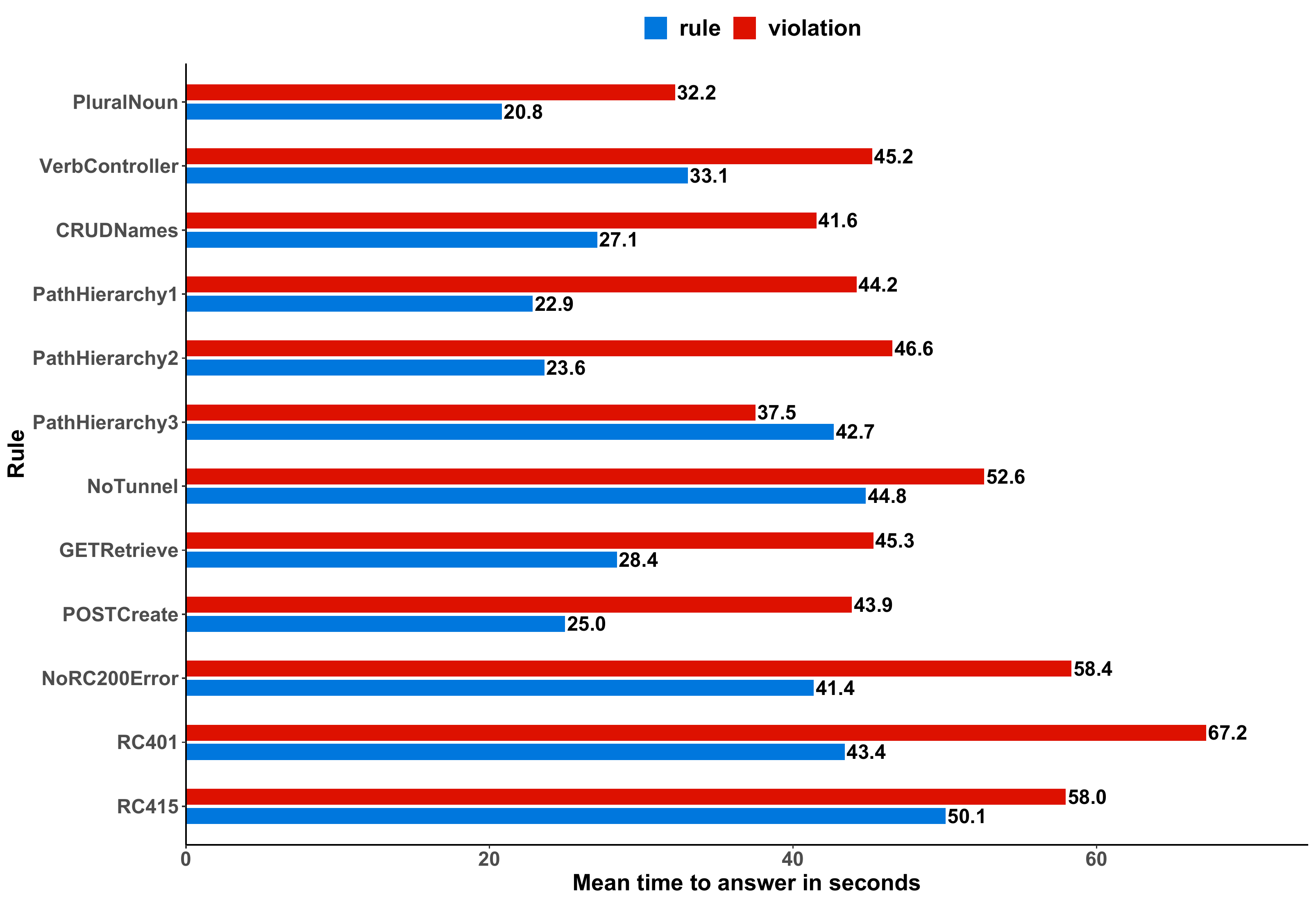}
    \vspace{-0.2cm}
    \caption{Comparison of duration between treatments \textit{rule} and \textit{violation} (lower is better)}
    \label{fig:barplotTime}
\end{figure}

\begin{figure}[H]
    \centering
    \includegraphics[width=\textwidth, trim={0.2cm 0.3cm 1.4cm 0.4cm}, clip]{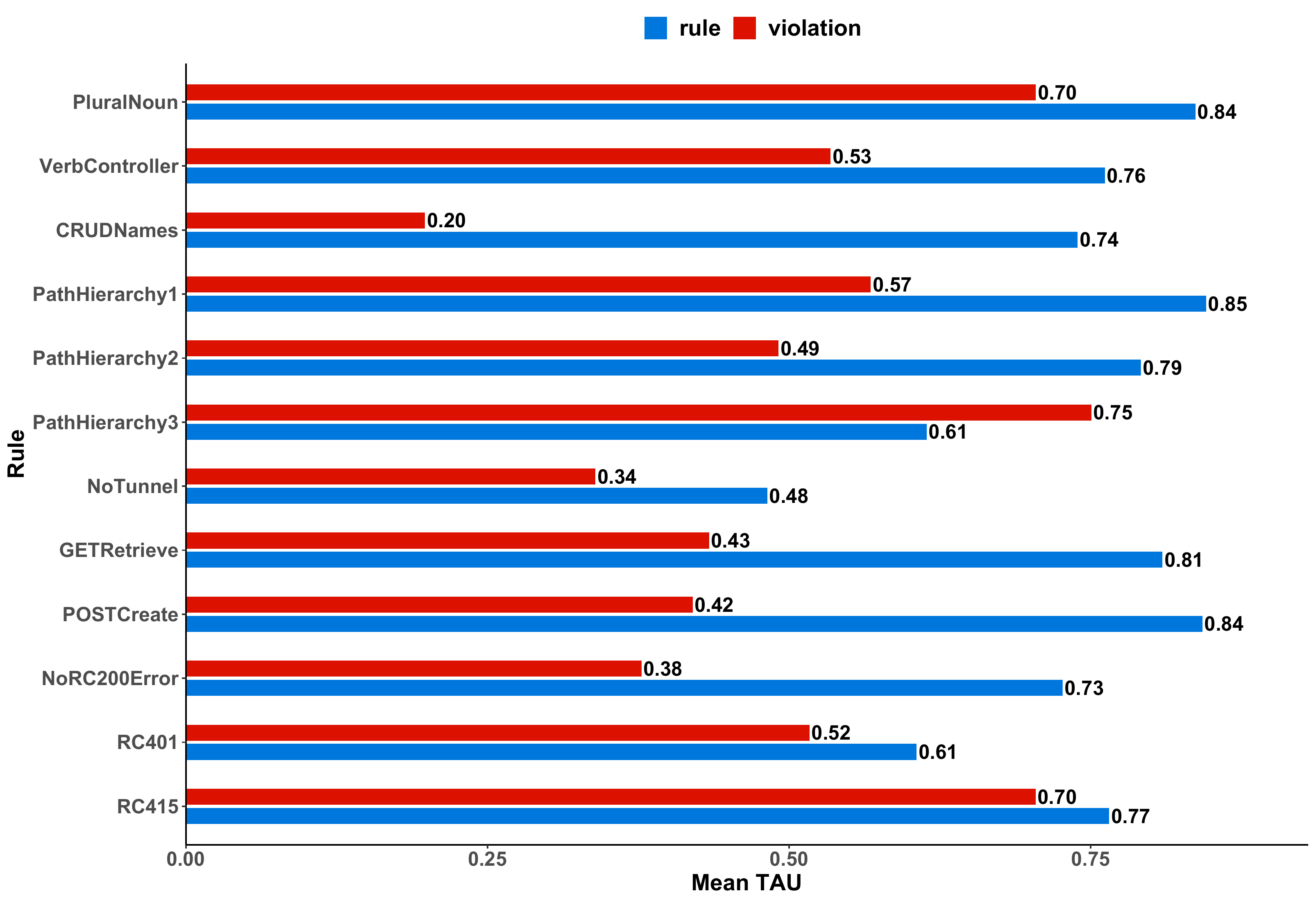}
    \vspace{-0.2cm}
    \caption{Comparison of TAU between treatments \textit{rule} and \textit{violation} (higher is better)}
    \label{fig:barplotTAU}
\end{figure}

For our central research question, we wanted to analyze which design rules have a significant impact on understandability, as measured via comprehension tasks and TAU.
We visualize the results for the percentage of correct answers (Fig.~\ref{fig:barplotPercent}), the required time to answer (Fig.~\ref{fig:barplotTime}), and TAU (Fig.~\ref{fig:barplotTAU}) per individual rule and treatment.
For a more detailed comparison, Tables~\ref{tab:resultsRQ1}, \ref{tab:extendedResultsTAU}, and~\ref{tab:extendedResultsTime} in the appendix list the descriptive statistics for this per task.

For 11 of the 12 tasks, participants with the \textit{rule} version performed better than participants with the \textit{violation} version, i.e., mean TAU was higher for \textit{rule}.
Correctness was often the deciding factor, e.g., for the tasks \texttt{CRUDNames} (96\% vs. 38\%), \texttt{GETRetrieve} (100\% vs. 60\%), or \texttt{NoRC200Error} (98\% vs. 58\%).
In other cases, correctness was much closer between the two treatments, but participants with \textit{violation} required more time.
This is, e.g., visible for the tasks \texttt{PluralNoun} (100\% vs. 94\%, but 20.82 s vs. 32.23 s) or \texttt{RC401} (75\% vs. 70\%, but 43.42 s vs. 67.21 s).
The only surprising exception was the task \texttt{PathHierarchy3}, where participants with \textit{violation} performed notably better (mean TAU of 0.6141 vs. 0.7505).

\begin{figure}[H]
    \centering
    \includegraphics[width=\textwidth]{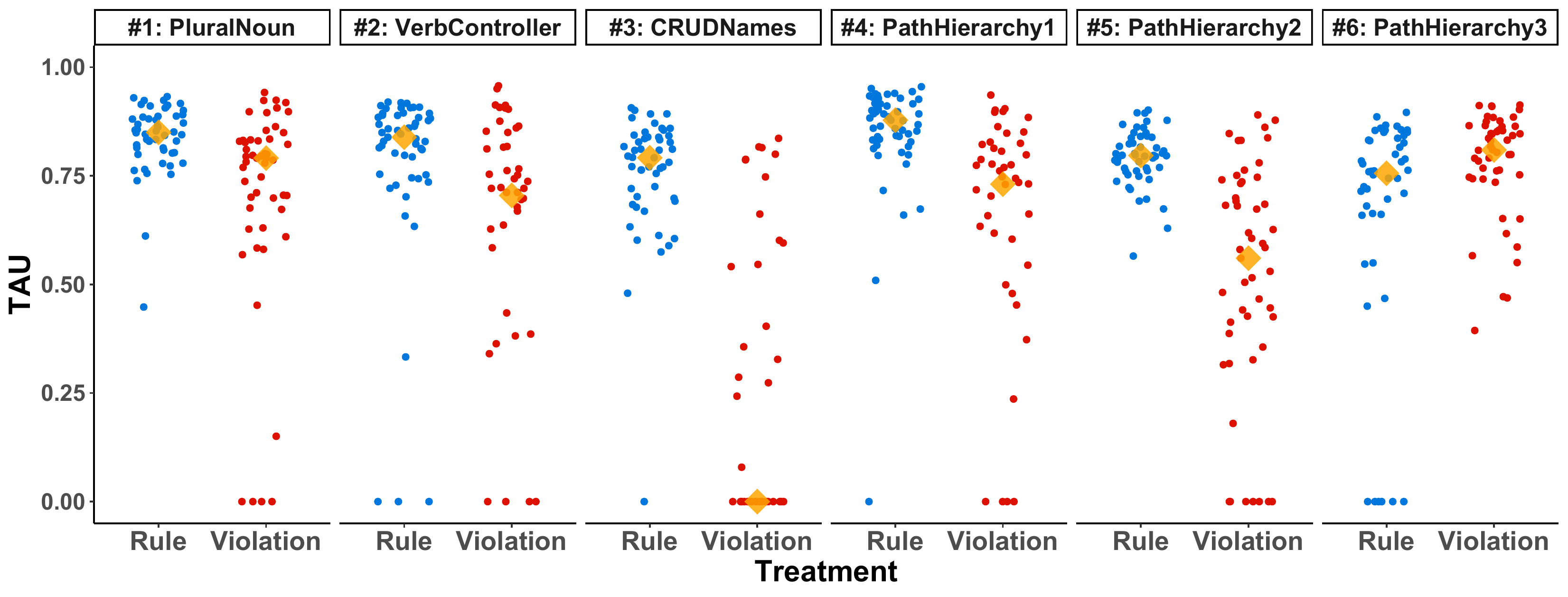}
    \vspace{-0.3cm}
    \caption{TAU distributions for the categories \textit{URI Design} (1-3) and \textit{Hierarchy Design} (4-6), median values are displayed as orange diamonds}
    \label{fig:stripplotTAU1}
\end{figure}

To further visualize the experiment performance, we created strip plots of TAU for all rules, which makes it easier to understand its unusual distribution and to compare the two treatments.
Fig.~\ref{fig:stripplotTAU1} shows the TAU distributions for the rules in the categories \textit{URI Design} and \textit{Hierarchy Design}.
Incorrect answers are displayed as dots at the bottom (TAU = 0), with the median value being displayed as an orange diamond.
For all three \textit{URI Design} tasks (1-3), it immediately becomes apparent that \textit{violation} (red) performed worse, i.e., the median of \textit{violation} is below \textit{rule} in all cases.
This difference is especially large for the task \texttt{CRUDNames}, which constitutes the worst performance for \textit{violation}. 
Additionally, the values are more spread out for \textit{violation} in all three tasks.

For the rules of the \textit{Hierarchy Design} category (4-6), we see a similarly improved performance for \textit{rule} in the case of \texttt{PathHierarchy1} and \texttt{PathHierachy2}.
However, the exception is \texttt{PathHierarchy3}, which shows a visibly better performance and less spread for \textit{violation}.
A potential explanation could be that the general difference in path length might have been the deciding factor, i.e., \\
\texttt{POST /customers/\{customerId\}/environments/\{environmentId\}/servers}\\
vs. the much simpler \textit{violation} \texttt{POST /servers}.
Even though following the rule provides richer details about the resource hierarchies (\enquote{there are customers, who have IT environments like \textit{Production}, in which servers are placed}), it seems to make the endpoint purpose more difficult to understand, in this case that a new server is created.
Additionally, from the five response options, three were about creating a new resource, which fits to the \texttt{POST} of the shown endpoint.
However, it was probably clearer to exclude the options \enquote{Create multiple new servers...} and \enquote{Create an environment...} for \textit{violation}, as \enquote{environment} did not occur in the path.
The needed time was not that different ($42.69$ s vs. $37.51$ s), but only 83\% answered correctly for \textit{rule}, while 96\% did the same for \textit{violation}.
Extrapolating from the other \texttt{PathHierarchy} rules (1 and 2), we might state that using a path hierarchy correctly, i.e., from left to right with sequential pairs of \texttt{\{collectionName\}/\{id\}}, is significantly better than using it \textit{incorrectly}, but not automatically better than \textit{not using it at all}.
However, more research is needed to confirm this, e.g., by analyzing other variants of \texttt{PathHierarchy3}, e.g., with \texttt{GET} instead of \texttt{POST}, and identifying potential thresholds below which the path length might become irrelevant.

\begin{figure}[H]
    \centering
    \includegraphics[width=\textwidth]{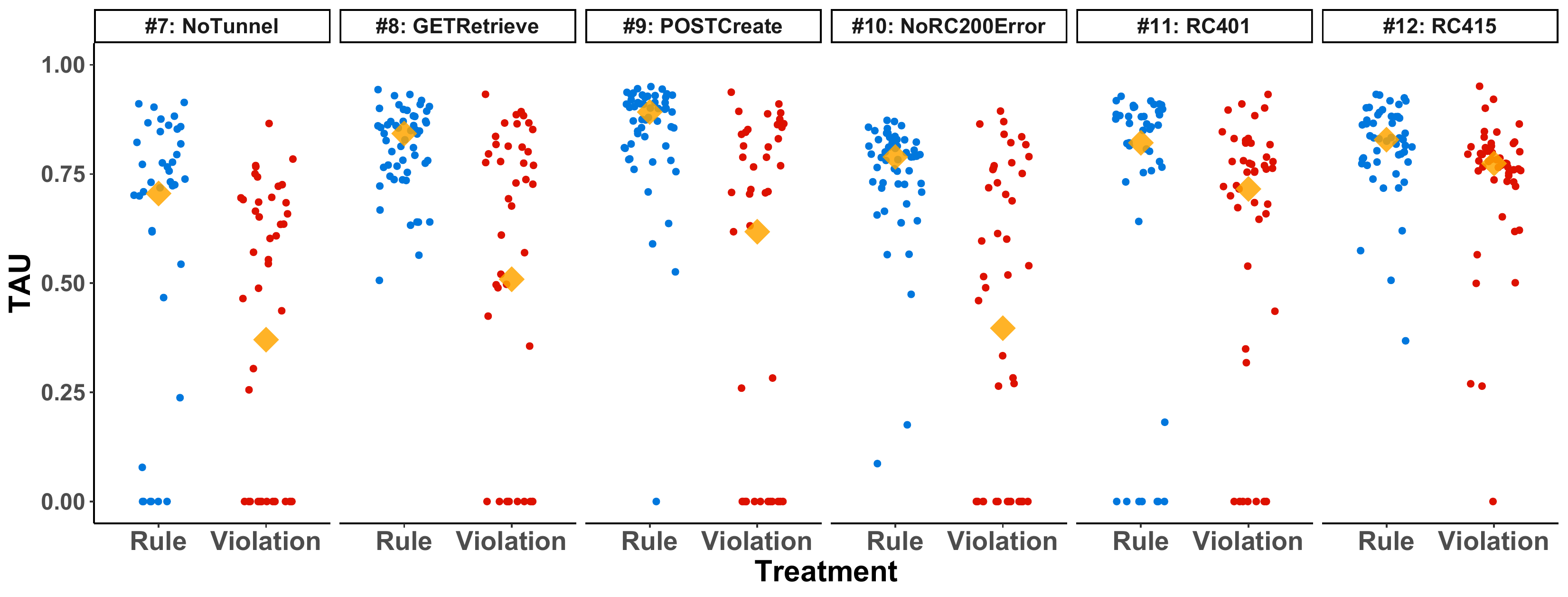}
    \vspace{-0.3cm}
    \caption{TAU distributions for the categories \textit{Request Methods} and \textit{HTTP Status Codes}, median values are displayed as orange diamonds}
    \label{fig:stripplotTAU2}
\end{figure}

Fig.~\ref{fig:stripplotTAU2} presents the TAU distributions for the categories \textit{Request Methods} and \textit{HTTP Status Codes}.
For \textit{Request Methods} (7-9), the \textit{rule} version shows again better performance in all three tasks, with substantial distance between treatment medians.
However, in the \texttt{NoTunnel} task, the spread of TAU values is more similar between \textit{rule} and \textit{treatment} than for other tasks.
Overall, \texttt{NoTunnel} was the worst performance for \textit{rule}, with only 66\% answering it correctly. 
One explanation may be that several practitioners may not see it as a violation to use \texttt{PUT} to create a new resource.

In the final \textit{HTTP Status Codes} category (10-12), the performance of \textit{rule} is also better in all three cases, however, to varying degrees.
For \texttt{NoRC200Error}, the differences are substantial, but for \texttt{RC401} and especially \texttt{RC415}, the median values are not as far apart.
In the case of \texttt{RC401}, the spread in TAU values is also much more similar between the two treatments.
This was the second-worst performance for \textit{rule}, with only 75\% providing the correct answer.

\begin{figure}[H]
    \centering
    \includegraphics[width=\textwidth]{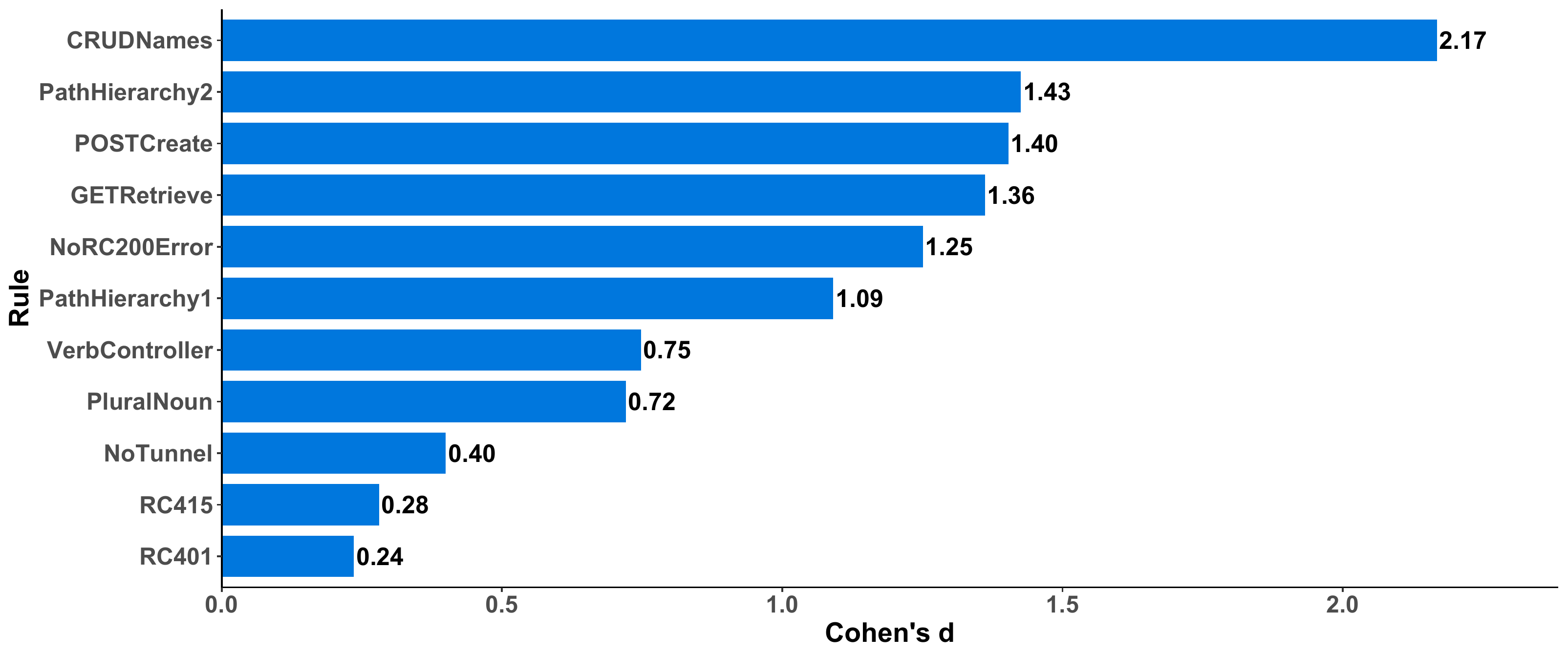}
    \vspace{-0.2cm}
    \caption{Effect sizes for the 11 significant differences between TAU for treatments \textit{rule} and \textit{violation}, ordered by Cohen's $d$, insignificant \texttt{PathHierarchy3} not included}
    \label{fig:cohensdRQ1}
\end{figure}

To verify if the visually identified differences between \textit{rule} and \textit{violation} were statistically significant, we continued with hypothesis testing.
Despite the applied Holm-Bonferroni correction, \textbf{11 of our 12 hypothesis tests produced significant results}, i.e., 11 of 12 rules had significant impact on understandability.
The only outlier was \texttt{PathHierarchy3}, where \textit{violation} had performed better.
Additionally, we calculated the effect size for the significant tests via Cohen's $d$.
We visualize the $d$ values in Fig.~\ref{fig:cohensdRQ1}.
For a more detailed comparison of the test results, please refer to Table~\ref{tab:testResultsRQ1} in the appendix.

\vspace{-0.2cm}

\vspace{-0.3cm}
The rules \texttt{NoTunnel}, \texttt{RC401}, and \texttt{RC415} produced a \textit{small} effect ($0.2<d<0.5$) and \texttt{PluralNoun} and \texttt{VerbController} a \textit{medium} one ($0.5<d<0.8$).
The remaining six rules showed even stronger effects, namely a \textit{large} one for \texttt{PathHierarchy1} ($0.8<d<1.2$), a \textit{very large} one for \texttt{NoRC200Error}, \texttt{GETRetrieve}, \texttt{POSTCreate}, and \texttt{PathHierarchy2} ($1.2<d<2.0$), and even a \textit{huge} one for \texttt{CRUDNames} ($d>2.0$).
It is difficult to determine the most impactful category of rules, as the six rules with a Cohen's $d>1.0$ cover all four categories.
For \textit{URI Design}, violating \texttt{CRUDNames} had by far the largest effect ($d=2.17$), but \texttt{VerbController} ($d=0.75$) and \texttt{PluralNoun} ($d=0.72$) were in the bottom half.
The categories \textit{Request Methods} and \textit{Hierarchy Design} also had two impactful rules, but \texttt{NoTunnel} produced only a small effect and \texttt{PathHierarchy3} none at all.
Lastly, \textit{HTTP Status Codes} was the least impactful category.
Even though \texttt{NoRC200Error} had a Cohen's $d=1.25$, both \texttt{RC401} and \texttt{RC415} resulted in small effects.

\begin{tcolorbox}
    \textbf{Results for RQ1:}
    For 11 of the 12 tasks (\texttt{PathHierarchy3} being the exception), adhering to the design rule resulted in significantly better comprehension performance.
    Effect sizes were between $0.24$ and $2.17$, with 6 rules resulting in a Cohen's $d>1.0$ (\textit{large} up to \textit{huge} effects).
    Violating the rules \texttt{CRUDNames}, \texttt{PathHierarchy2}, and \texttt{POSTCreate} had the strongest negative impact on understandability.
\end{tcolorbox}

\subsection{Impact on Perceived Difficulty (RQ2)}
After showing that the majority of design rules led to significantly better comprehension performance, we analyzed if the results were similar for the subjective \textit{perceived} understandability.
Table~\ref{tab:resultsRQ2} summarizes the results for the perceived difficulty ratings and sets them in relation to the corresponding TAU values.
The mean difficulty rating is lower for \textit{rule} in all 12 tasks, while the median difficulty rating is only lower for 10, the exceptions being \texttt{RC415} and again \texttt{PathHierarchy3} (median rating of 2 for both treatments).
In four tasks, namely \texttt{VerbController}, \texttt{CRUDNames}, \texttt{GETRetrieve}, and \texttt{NoRC200Error}, the difference in median rating is 1 point, while it is even higher for the remaining six tasks (between 1.5 and 3 points).
To further analyze differences, we visualized the results with a comparative bar plot of the ratings 1 (\textit{very easy}) and 2 (\textit{easy}) in Fig.~\ref{fig:customLikertPlot}.
For a full Likert plot, please refer to Fig.~\ref{fig:likertPlots} in the appendix.

\begin{table}[H]
    \caption{Descriptive statistics for RQ2, perceived difficulty rating ranging from 1 (\textit{very easy}) to 5 (\textit{very difficult}), mean TAU provided for comparison}
	\label{tab:resultsRQ2}
	\centering
    \begin{tabular}{lllllll}
        \multirow{2}{*}{Task} &
        \multicolumn{2}{c}{Median difficulty} &
        \multicolumn{2}{c}{Mean difficulty} &
        \multicolumn{2}{c}{Mean TAU} \\
        & \textit{rule} & \textit{violation} & \textit{rule} & \textit{violation} & \textit{rule} & \textit{violation} \\
        \hline
        \hline
        \texttt{PluralNoun} & 1 & 2.5 & 1.25 & 2.40 & 0.8369 & 0.7046 \\
        \texttt{VerbController} & 2 & 3 & 2.13 & 2.50 & 0.7617 & 0.5345 \\
        \texttt{CRUDNames} & 2 & 3 & 1.73 & 3.36 & 0.7392 & 0.1982 \\
        \texttt{PathHierarchy1} & 1 & 3 & 1.34 & 2.94 & 0.8457 & 0.5678 \\
        \texttt{PathHierarchy2} & 1 & 4 & 1.23 & 3.36 & 0.7916 & 0.4913 \\
        \texttt{PathHierarchy3} & 2 & 2 & 2.31 & 2.42 & 0.6141 & 0.7505 \\
        \texttt{NoTunnel} & 2 & 4 & 2.34 & 3.27 & 0.4820 & 0.3396 \\
        \texttt{GETRetrieve} & 2 & 3 & 1.83 & 2.77 & 0.8095 & 0.4337 \\
        \texttt{POSTCreate} & 1 & 3 & 1.45 & 2.75 & 0.8427& 0.4201 \\
        \texttt{NoRC200Error} & 2 & 3 & 1.75 & 2.80 & 0.7268 & 0.3776 \\
        \texttt{RC401} & 2 & 4 & 1.80 & 3.45 & 0.6058 & 0.5170 \\
        \texttt{RC415} & 2 & 2 & 2.00 & 2.35 & 0.7654 & 0.7044 \\
        \hline
        \hline
    \end{tabular}
\end{table}

The difference between \textit{rule} and \textit{violation} becomes apparent for many of the tasks here, e.g., for \texttt{CRUDNames}, \texttt{PathHierarchy1}, \texttt{PathHierarchy2}, and \texttt{RC401}.
However, in some tasks, the treatment ratings also appear to be decently close to each other, despite the median rating being different, e.g., for \texttt{VerbController} or \texttt{GETRetrieve}.
In general, the \textit{violations} perceived as the least difficult to understand were \texttt{PathHierarchy3}, \texttt{RC415}, and \texttt{PluralNoun}.

\begin{figure}[h]
    \centering
    \includegraphics[width=\textwidth]{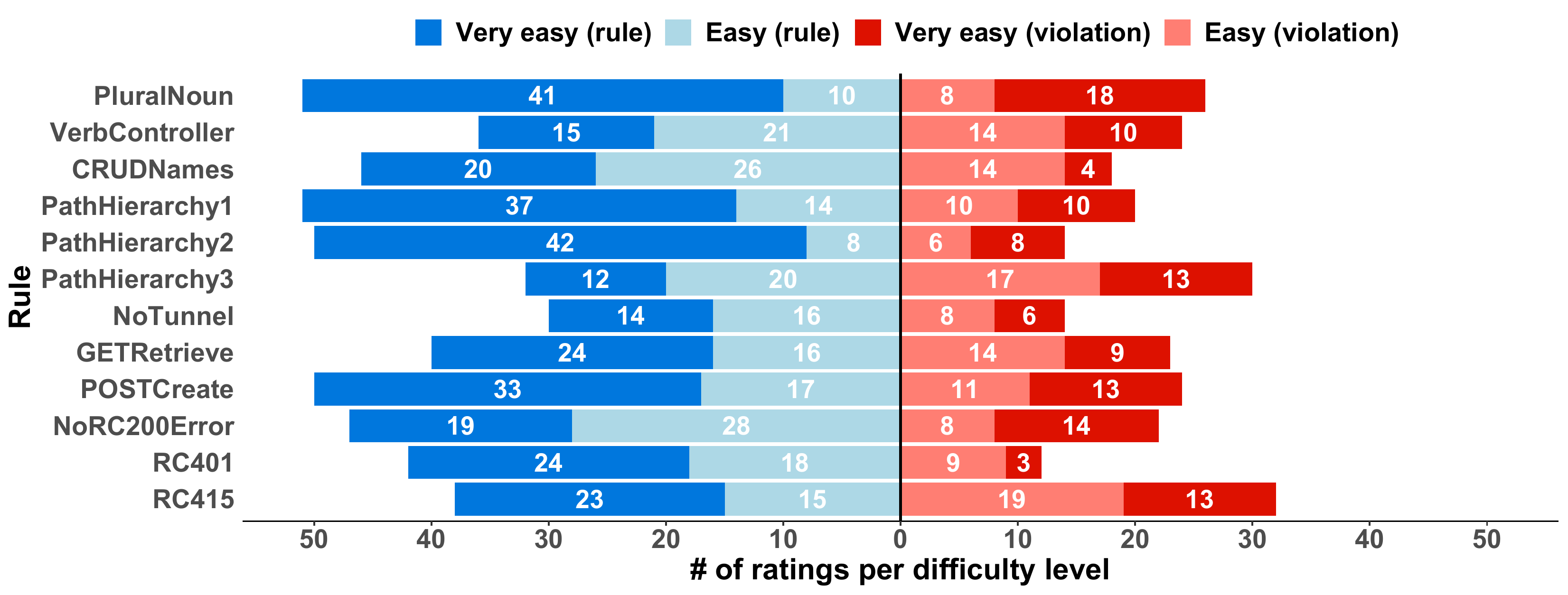}
    \caption{Bar plots of perceived difficulty (RQ2) for the ratings 1 (\textit{very easy}) and 2 (\textit{easy}), \textit{rule} ratings on the left, \textit{violation} ratings on the right}
    \label{fig:customLikertPlot}
\end{figure}

\begin{figure}[H]
    \centering
    \includegraphics[width=\textwidth]{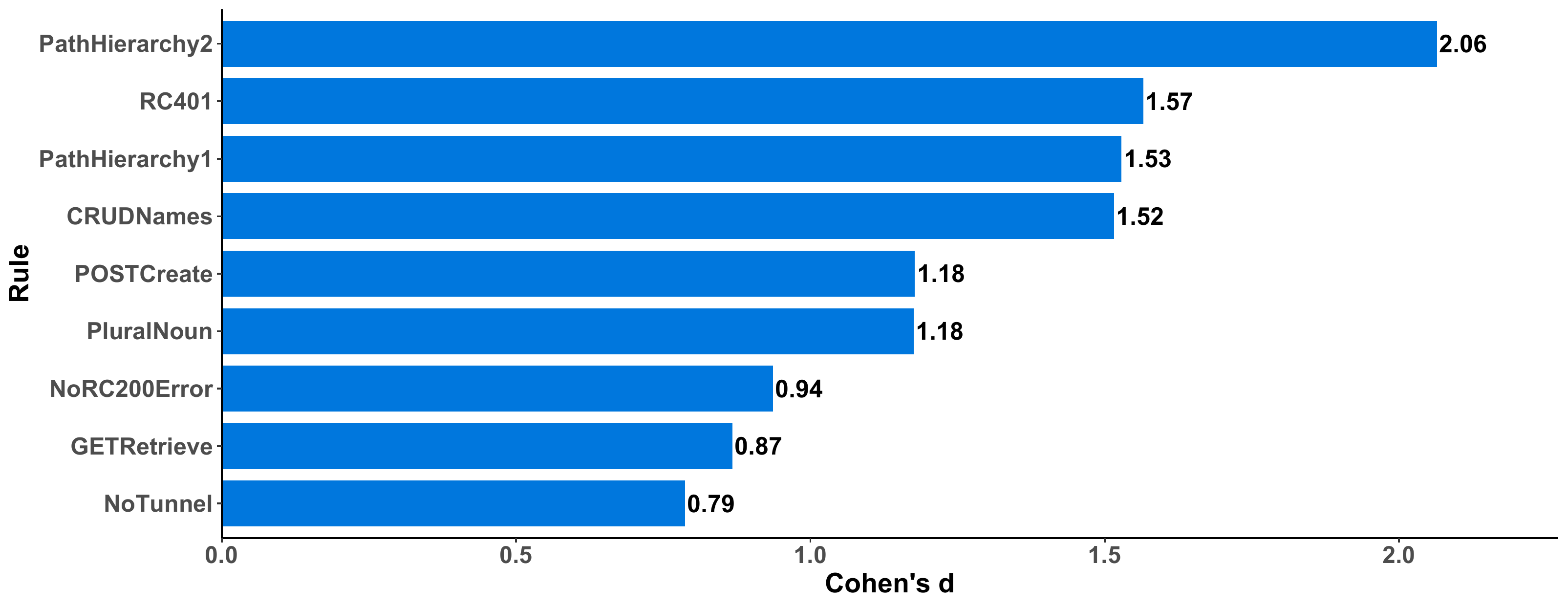}
    \caption{Effect sizes for the 9 significant differences between perceived difficulty for treatments \textit{rule} and \textit{violation}, ordered by Cohen's $d$, insignificant \texttt{PathHierarchy3}, \texttt{RC415}, and \texttt{VerbController} not included}
    \label{fig:cohensdRQ2}
\end{figure}

To confirm if the differences between \textit{rule} and \textit{violation} are significant, we again applied hypothesis testing.
This time, we only found \textbf{significant differences for 9 of the 12 tasks} in the perception of \textit{rule} and \textit{violation}.
In addition to \texttt{PathHierarchy3} and \texttt{RC415} with equal medians, the difference between treatments for \texttt{VerbController} was also not significant.
We visualize the values for Cohen's $d$ in Fig.~\ref{fig:cohensdRQ2}.
For a more detailed comparison, please refer to Table~\ref{tab:testResultsRQ2} in the appendix.
The effect sizes for the significant tasks ranged from $0.79$ (\textit{medium}) to $2.06$ (\textit{huge}), with 8 rules producing a Cohen's $d>0.8$ (\textit{large} and higher).
Violating the rules \texttt{PathHierarchy2}, \texttt{RC401}, \texttt{PathHierarchy1}, and \texttt{CRUDNames} had the strongest impact on difficulty perception.

Especially for \texttt{PathHierarchy2} with Cohen's $d=2.06$ (\textit{huge}), we see that the constructed URI for \textit{violation} (\texttt{GET /employees/companies/\{companyId\}}) was a very extreme case leading to much confusion.
Mixing the URI path segments in this way seems to have made it much more difficult for participants to identify the cardinalities of the original domain model (all employees belonging to a specific company).
For \texttt{PathHierachy3}, the missing effect was to be expected and complementary to the insignificant results for TAU.
However, \texttt{VerbController} and \texttt{RC415} are not as easy to explain, as both of them had a significant impact on the actual understandability.
While \textit{violation} was perceived as slightly more difficult than \textit{rule} in both tasks, this difference was not statistically significant.
For \texttt{RC415}, this might be explained by the fairly small effect size for the TAU difference ($d=0.28$) and the similar levels of correctness per treatment.
If participants only needed a bit more time but were still fairly certain to have found the correct answer, they might not have directly associated this task with a high difficulty. 
For \texttt{VerbController}, however, this explanation is not applicable, as both correctness and time differed per treatment, and it produced a medium effect ($d=0.75$).
This makes it an especially dangerous rule to violate because it has considerable impact, but many people might not notice that something is ambiguous or unclear with the endpoint.

\begin{figure}[H]
    \centering
    \includegraphics[width=\textwidth]{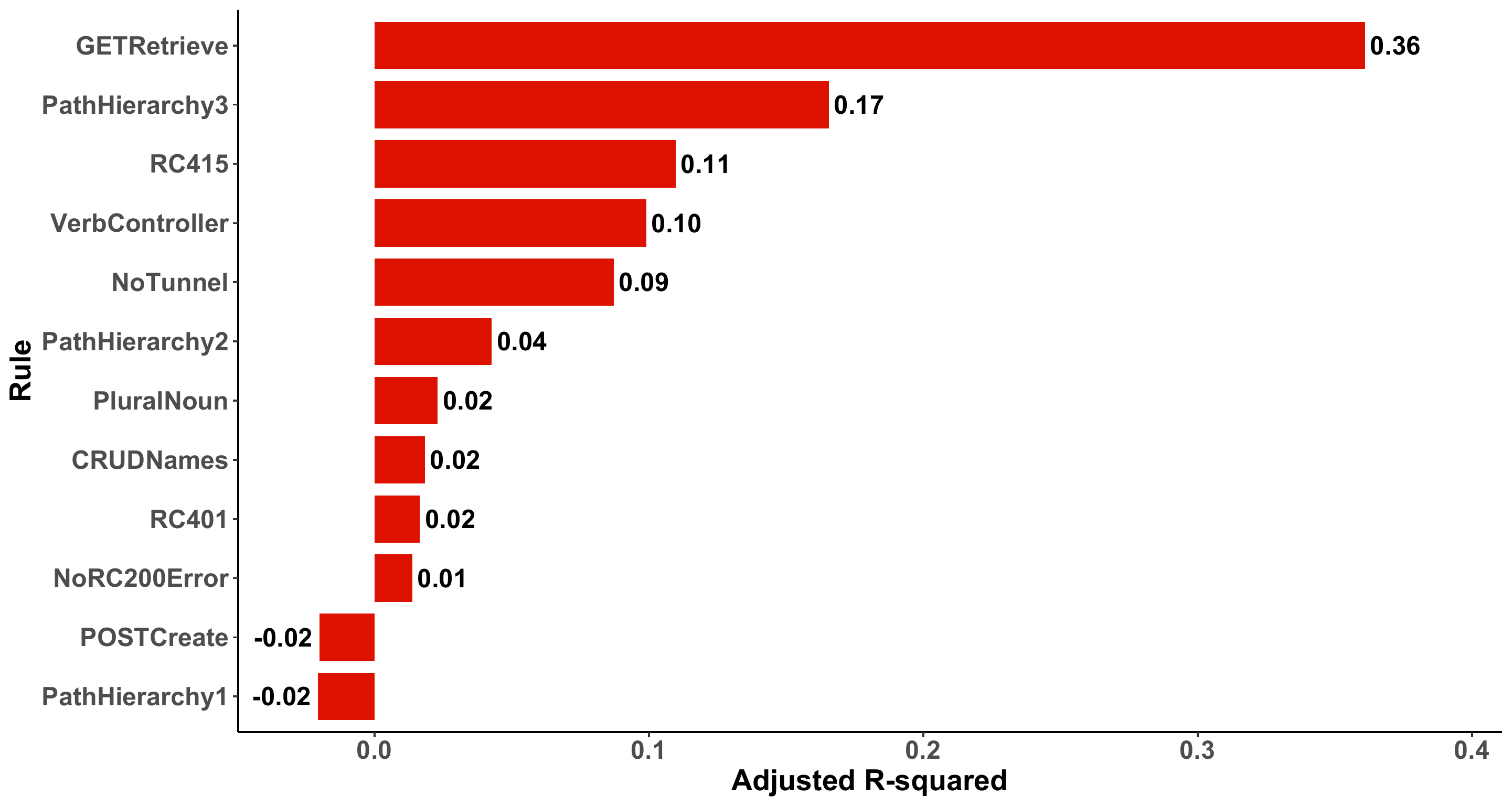}
    \caption{Adjusted $R^2$ values for the regression between TAU and perceived difficulty ratings for \textit{violation}, ordered by adjusted $R^2$, higher is better}
    \label{fig:adjustedR2}
\end{figure}

Other rule violations that may be especially critical can be identified by analyzing the correlations between TAU and perceived difficulty for the \textit{violation} treatment.
Ideally, we would like to have significant negative correlations between the two dependent variables for all \textit{violation} API snippets, i.e., the worse the comprehension performance of a rule violation, the higher the perceived difficulty ratings should be.
Rule violations where this is not the case may indicate that participants felt confident in their performance despite answering incorrectly or slowly because they unknowingly misunderstood the API snippet.
For nine tasks, there was \textit{no} significant negative correlation between TAU and perceived difficulty, the exceptions being \texttt{PathHierarchy3}, \texttt{VerbController}, and \texttt{GETRetrieve}.
For the detailed correlation results, please refer to Table~\ref{tab:correlationsRQ2} in the appendix.
As a measure of explanatory power, we visualize the adjusted $R^2$ values for the regression between TAU and the perceived difficulty ratings for all \textit{violation} API snippets in Fig.~\ref{fig:adjustedR2}.
For many of them, the values are close to zero, meaning that the perceived difficulty ratings cannot explain any variation in TAU.
Among them are also rule violations that have a substantial impact on understandability, like \texttt{CRUDNames}, \texttt{PathHierarchy2}, or \texttt{POSTCreate}, which makes these rule violations especially problematic.

\begin{tcolorbox}
    \textbf{Results for RQ2:}
    For 9 of the 12 tasks (except for \texttt{PathHierarchy3}, \texttt{RC415}, and \texttt{VerbController}), adhering to the API design rule resulted in significantly smaller perceived difficulty ratings than violating it.
    Effect sizes were between $0.79$ and $2.06$, with 6 rules resulting in a Cohen's $d>1.0$ (\textit{large} up to \textit{huge} effects).
    Additionally, there was no significant relationship between perceived difficulty and TAU for many \textit{violation} tasks, making these rule violations especially dangerous.
\end{tcolorbox}

\subsection{Relationships with Demographic Attributes (RQ3)}
For RQ3, we analyzed if there were any relationships between the dependent variables (TAU and the perceived difficulty ratings) and demographic attributes of our participants.
We explored this question separately for each treatment (\textit{rule} and \textit{violation}), and compared the results.
Studied predictors were being from Germany, being from academia (vs. industry), being a student, being an API developer / designer, years of professional experience with REST, having knowledge of the Richardson maturity model, and the preferred minimal maturity level.
Even though this was an exploratory RQ, we used Holm-Bonferroni adjusted p-values and a significance level of $\alpha=0.05$.

Overall, we did not find many significant relationships, and no strong ones at all.
Years of experience with REST had a small positive correlation with TAU for \textit{rule} (Kendall's $\tau=0.1956$, $p=0.0311$), i.e., for the API snippets adhering to the rules, participants with more experience had a slight tendency to perform better.
However, this was not the case for API snippets violating the rules.
Similarly, participants knowing the Richardson maturity model also tended to perform slightly better for \textit{rule}, with a Kendall's $\tau=0.1941$.
However, after adjusting the original p-value ($p=0.0158$), this correlation was no longer significant ($p=0.0791$).
The deciding factor here should be years of experience, though, as it was also positively correlated with knowing the Richardson maturity model in our sample (Kendall's $\tau=0.3760$, $p<0.001$).
Conversely, knowledge of the Richardson maturity model was also positively correlated with the perceived difficulty ratings for \textit{violation}, i.e., if participants knew about this model, they tended to rate the API snippets violating the rules as slightly more difficult to understand (Kendall's $\tau=0.3035$, $p=0.0014$).
This correlation was absent for \textit{rule}.
All other demographic attributes did not produce any significant relationships.

While the identified correlations were small, they still seem to highlight differences between the two treatments that may provide an explanation.
Experience and knowledge about REST is only linked to better experiment performance if no design rules are violated.
If rules are violated, it does not matter much if people are more experienced: their performance still suffers.
However, people who know advanced REST-related concepts are at least more likely to \textit{notice} that something is wrong with the rule-violating API snippets, even though this does not help to understand them better.
This theory also seems to be supported by our results of trying to build linear regression models to predict TAU per treatment based on the demographic attributes.
While both models (\textit{rule} and \textit{violation}) were unable to provide reliable predictions for the majority of our sample, the model for \textit{violation} performed considerably worse.
The model for \textit{rule} was able to explain a decent percentage of variability (adjusted $R^2=0.2171$, $p=0.0618$), while the one for \textit{violation} had no explanatory power at all (adjusted $R^2=-0.1792$, $p=0.9674$).

\begin{tcolorbox}
    \textbf{Results for RQ3:}
    Participants with more REST-related experience and knowledge of the Richardson maturity model performed slightly better within \textit{rule}, but not within \textit{violation}.
    Conversely, people knowing the Richardson maturity model perceived the \textit{violation} API snippets as slightly more difficult to comprehend.
    This seems to indicate that violating design rules decreases understandability regardless of demographic attributes like experience.
\end{tcolorbox}

\section{Threats to Validity}
This section describes how we tried to mitigate potential threats to validity, and which threats and limitations to our results remain.
We discuss these issues mainly through the perspectives provided by \citet{Wohlin2012}.

\textbf{Construct validity} is concerned with relating the experiment and especially its collected measures to the studied concepts.
This includes whether our dependent variables were adequate representations of the constructs.
Our measure for understandability, namely TAU as a combination of correctness and time~\citep{Scalabrino2021}, has been used in several studies before.
It provides the advantage of increasing the information density in a single statistical test.
However, TAU's trade-off between time and correctness is not ideal when correctness is binary.
TAU therefore definitely has limitations, but looking at time and correctness separately to interpret the results allows reducing these threats.
Regarding the perceived understandability, it is an accepted practice to use ordinal scales that are balanced and symmetrical for subjective ratings.

To avoid \textit{hypothesis guessing}~\citep{Wohlin2012}, participants were told that the goal of the experiment was to analyze the understandability of different Web API designs.
We see it as very unlikely that participants figured out the true goal of the experiment and then deliberately gave worse responses for \textit{violation} tasks.
Lastly, it is possible that our experiment may be slightly impacted by a \textit{mono-operation bias}~\citep{Wohlin2012}, i.e., the underrepresentation of the construct by focusing on a single treatment.
While we used different categories with several rules each in the experiment, we still only compared one exemplary rule implementation to its complementary violation in isolation.
We did not combine several rules or involved other RESTful API concepts, which could have led to a richer theory.

\textbf{Internal validity} can suffer from threats that may impact the dependent variables without the researchers' knowledge, i.e., \textit{confounders}~\citep{Wohlin2012}.
In general, a crossover design is fairly robust against many confounders by reducing the impact of inter-participant differences, e.g., large expertise or experience differences between participants.
The relatively short experiment duration of 10-15 minutes caused by our between-subjects characteristic also made most \textit{history}~\citep{Wohlin2012} or \textit{maturation}~\citep{Wohlin2012} effects unlikely, e.g., noticeable changes in performance due to tiredness or boredom.
However, due to the similar task structures, learning effects are very likely.
Randomization of the task order could have mitigated this, but it could also have led to suboptimal sequences of treatments and task categories, which could have introduced harmful carryover effects.
Our fixed sequences at least guaranteed that both treatments for the same task appeared at the same position, thereby ensuring a similar level of maturation per experiment object.
Additionally, randomization was used to assign participants to the two sequences and to display the different answers per comprehension question.

This study was conducted as an online experiment, with considerably less control over the experiment environment.
While \textit{random irrelevancies in the experimental setting}~\citep{Wohlin2012} might have occurred in some cases, such as reduced concentration due to loud noise or an interruption by colleagues or family members, this potential increase in variance still did not impact our hypothesis tests.
Moreover, we also had no means to prevent people from participating several times, as responses were anonymous.
Due to no reasonable incentive for this, we deem this threat as very unlikely.

\textbf{External validity} is the extent to which the results are generalizable to other settings or parts of the population.
Our sample was fairly diverse, with participants with different levels of experience from both industry and academia.
With 105 participants, our sample size was also decently large when compared to many other software engineering experiments, even though we need to remember that only half of our participants worked on each treatment per individual rule due to our hybrid design.
While we reached statistical significance for most hypotheses, a larger sample would have improved the generalizability of the results even more.
Nonetheless, we see it as unlikely that the \textit{interaction of selection and treatment} threat~\citep{Wohlin2012} could impact our results.
Even though most of our sample was located in Germany (67\%), we do not believe that country-specific differences might have a noticeable influence on the results.
To combat the threat \textit{interaction of setting and treatment}~\citep{Wohlin2012}, we ensured using realistic API concepts and violations inspired by real-world examples.
The Swagger editor is also a very popular tool in the area of Web APIs.

Finally, we need to emphasize that the understandability of Web APIs was tested in a consciously constructed setting, without a hosted implementation for manual exploration or additional API documentation.
Participants could only rely on the provided API snippets in the visual notation from the Swagger editor.
This is obviously different from a real-world environment, e.g., in industry, where a software engineer trying to use a Web API may have access to the API documentation or even a running API instance for manual testing.
However, in many cases, access to more documentation or the API itself is also not available in the real world.
Additionally, using these artifacts also requires additional time, i.e., being able to understand the purpose of an endpoint without having to consult other materials is still preferable.
Moreover, it is also plausible that some rules might even have a stronger effect in a real-world setting, e.g., \texttt{RC401} and \texttt{RC415}: getting access to HTTP header information requires a lot more effort there, while we conveniently presented the headers in the experiment.
All in all, we believe that our results are decently transferrable to less controlled real-world environments.
To analyze the full degree of this generalization, follow-up research is necessary, e.g., based on repository mining or industrial case studies.

\section{Conclusion and Future Work}
In this paper, we presented the design and results of our controlled Web-based experiment on the understandability impact of 12 RESTful API design rules from \citet{Masse2011} and \citet{Richardson2007}.
In detail, we presented 12 Web API snippets to 105 participants, asked them comprehension questions about each snippet, and let them rate the perceived understandability.
For 11 of 12 rules, we identified a significant negative impact on understandability for the \textit{violation} treatment.
Effect sizes ranged from \textit{small} to \textit{huge}, with Cohen's~$d$ between $d=0.24$ and $d=2.17$.
Furthermore, our participants also rated 9 of 12 rule violations as significantly more difficult to understand.

All in all, our results indicate that violating commonly accepted design rules for RESTful APIs has a negative impact on understandability, regardless of REST-related experience or other demographic factors.
For several rule violations, we could also show that they are prone to misinterpretation and misunderstandings, making them especially dangerous.
Practitioners should therefore respect these rules during the design of Web APIs, as understandability is linked with important quality attributes like maintainability and usability.
This becomes especially important for publicly available APIs that are meant to be used by many external people.
Providing comprehensive API documentation may partly mitigate understandability problems caused by rule violations, but is still an insufficient solution.

In the future, additional experiments should try to replicate these findings with other samples of the population or other violations of the same rules, and potentially extend the evidence to other rules not tested in our study.
To enable such studies, we publicly share our experiment materials\footnote{\url{https://doi.org/10.5281/zenodo.7381500}}.
Additionally, tool-supported approaches to automatically identify these rule violations will be helpful for practitioners trying to ensure the quality of their Web APIs.
Such tool support would also pave the way for large-scale studies of these rule violations in less controlled environments, such as software repository mining or industrial case studies.

\begin{acknowledgements}
We kindly thank all our experiment participants for their valuable time!
We also thank the experts participating in our pilot for their detailed feedback!
Lastly, we thank Dr. Daniel Graziotin (University of Stuttgart) and Dr. Sira Vegas (Universidad Politécnica de Madrid) for discussing the experiment design and its terminology with us.
\end{acknowledgements}

%
%

\bibliographystyle{spbasic}      
\bibliography{references}

\section*{Appendix}
\label{sec:appendix}

\begin{table}[H]
    \caption{Descriptive statistics for RQ1}
	\label{tab:resultsRQ1}
	\centering
    \begin{tabular}{lllllll}
        \multirow{2}{*}{Task} &
        \multicolumn{2}{c}{Mean TAU} &
        \multicolumn{2}{c}{Mean duration (s)} &
        \multicolumn{2}{c}{\# of correct answers} \\
        & \textit{rule} & \textit{violation} & \textit{rule} & \textit{violation} & \textit{rule} & \textit{violation} \\
        \hline
        \hline
        \texttt{PluralNoun} & 0.8369 & 0.7046 & 20.82 & 32.23 & 52 (100\%) & 49 (94\%) \\
        \texttt{VerbController} & 0.7617 & 0.5345 & 33.07 & 45.22 & 50 (94\%) & 38 (73\%) \\
        \texttt{CRUDNames} & 0.7392 & 0.1982 & 27.10 & 41.55 & 49 (96\%) & 20 (38\%) \\
        \texttt{PathHierarchy1} & 0.8457 & 0.5678 & 22.85 & 44.19 & 52 (98\%) & 40 (80\%) \\
        \texttt{PathHierarchy2} & 0.7916 & 0.4913 & 23.62 & 46.55 & 52 (100\%) & 44 (83\%) \\
        \texttt{PathHierarchy3} & 0.6141 & 0.7505 & 42.69 & 37.51 & 43 (83\%) & 50 (96\%) \\
        \texttt{NoTunnel} & 0.4820 & 0.3396 & 44.80 & 52.60 & 33 (66\%) & 28 (54\%) \\
        \texttt{GETRetrieve} & 0.8095 & 0.4337 & 28.40 & 45.29 & 53 (100\%) & 31 (60\%) \\
        \texttt{POSTCreate} & 0.8427& 0.4201 & 24.99 & 43.88 & 52 (98\%) & 29 (57\%) \\
        \texttt{NoRC200Error} & 0.7268 & 0.3776 & 41.37 & 58.35 & 52 (98\%) & 29 (58\%) \\
        \texttt{RC401} & 0.6058 & 0.5170 & 43.42 & 67.21 & 38 (75\%) & 37 (70\%) \\
        \texttt{RC415} & 0.7654 & 0.7044 & 50.05 & 57.97 & 51 (96\%) & 49 (94\%) \\
        \hline
        \hline
    \end{tabular}
\end{table}

\begin{table}[H]
    \caption{Extended descriptive statistics for TAU (RQ1)}
	\label{tab:extendedResultsTAU}
	\centering
    \begin{tabular}{lllllll}
        \multirow{2}{*}{Task} &
        \multicolumn{2}{c}{Min TAU} &
        \multicolumn{2}{c}{Max TAU} &
        \multicolumn{2}{c}{Variance of TAU} \\
        & \textit{rule} & \textit{violation} & \textit{rule} & \textit{violation} & \textit{rule} & \textit{violation} \\
        \hline
        \hline
        \texttt{PluralNoun} & 0.4479 & 0 & 0.9322 & 0.9418 & 0.0067 & 0.0606 \\
        \texttt{VerbController} & 0 & 0 & 0.9197 & 0.9572 & 0.0577 & 0.1276 \\
        \texttt{CRUDNames} & 0 & 0 & 0.9065 & 0.8362 & 0.0320 & 0.0913 \\
        \texttt{PathHierarchy1} & 0 & 0 & 0.9550 & 0.9359 & 0.0206 & 0.1118 \\
        \texttt{PathHierarchy2} & 0.5652 & 0 & 0.9013 & 0.8902 & 0.0043 & 0.0836 \\
        \texttt{PathHierarchy3} & 0 & 0 & 0.8959 & 0.9129 & 0.1008 & 0.0378 \\
        \texttt{NoTunnel} & 0 & 0 & 0.9139 & 0.8657 & 0.1436 & 0.1112 \\
        \texttt{GETRetrieve} & 0.5064 & 0 & 0.9433 & 0.9326 & 0.0095 & 0.1441 \\
        \texttt{POSTCreate} & 0 & 0 & 0.9502 & 0.9374 & 0.0219 & 0.1621 \\
        \texttt{NoRC200Error} & 0 & 0 & 0.8730 & 0.8941 & 0.0323 & 0.1263 \\
        \texttt{RC401} & 0 & 0 & 0.9280 & 0.9323 & 0.1533 & 0.1312 \\
        \texttt{RC415} & 0 & 0 & 0.9324 & 0.9511 & 0.0468 & 0.0477 \\
        \hline
        \hline
    \end{tabular}
\end{table}

\begin{table}[H]
    \caption{Extended descriptive statistics for response time (RQ1)}
	\label{tab:extendedResultsTime}
	\centering
    \begin{tabular}{lllllll}
        \multirow{2}{*}{Task} &
        \multicolumn{2}{c}{Min duration (s)} &
        \multicolumn{2}{c}{Max duration (s)} &
        \multicolumn{2}{c}{Variance of duration} \\
        & \textit{rule} & \textit{violation} & \textit{rule} & \textit{violation} & \textit{rule} & \textit{violation} \\
        \hline
        \hline
        \texttt{PluralNoun} & 8.65 & 7.43 & 70.48 & 127.66 & 109.92 & 510.73 \\
        \texttt{VerbController} & 11.55 & 7.38 & 172.27 & 113.62 & 671.21 & 660.98 \\
        \texttt{CRUDNames} & 11.05 & 12.59 & 61.47 & 118.14 & 130.32 & 660.62 \\
        \texttt{PathHierarchy1} & 7.45 & 6.53 & 81.19 & 165.46 & 181.03 & 934.83 \\
        \texttt{PathHierarchy2} & 11.19 & 12.45 & 49.28 & 113.35 & 54.90 & 495.81 \\
        \texttt{PathHierarchy3} & 16.05 & 14.20 & 162.99 & 140.22 & 638.31 & 607.81 \\
        \texttt{NoTunnel} & 13.40 & 17.02 & 155.65 & 134.77 & 971.09 & 874.44 \\
        \texttt{GETRetrieve} & 8.45 & 10.05 & 73.58 & 149.08 & 211.75 & 851.61 \\
        \texttt{POSTCreate} & 8.89 & 9.50 & 84.65 & 178.55 & 258.03 & 1146.90 \\
        \texttt{NoRC200Error} & 20.10 & 16.75 & 144.48 & 158.21 & 556.36 & 1083.60 \\
        \texttt{RC401} & 11.54 & 16.08 & 237.67 & 192.83 & 1588.84 & 1476.35 \\
        \texttt{RC415} & 15.38 & 11.13 & 227.46 & 167.35 & 1550.69 & 1001.35 \\
        \hline
        \hline
    \end{tabular}
\end{table}

\begin{table}[H]
	\caption{Hypothesis testing results for TAU (RQ1), Holm-Bonferroni adjusted p-values, $\alpha = 0.05$, sorted by effect size}
	\label{tab:testResultsRQ1}
	\centering
	\begin{tabular}{l r r r r}
        Task & Test statistic (U) & p-value & Cohen's $d$ & Accepted \\
        \hline
        \hline
        \texttt{CRUDNames} & $2434.0$ & $<0.001$ & $2.17$ & yes \\
        \texttt{PathHierarchy2} & $2359.0$ & $<0.001$ & $1.43$ & yes \\
        \texttt{POSTCreate} & $2281.5$ & $<0.001$ & $1.40$ & yes \\
        \texttt{GETRetrieve} & $2195.0$ & $<0.001$ & $1.36$ & yes \\
        \texttt{NoRC200Error} & $2077.5$ & $<0.001$ & $1.25$ & yes \\
        \texttt{PathHierarchy1} & $2225.5$ & $<0.001$ & $1.09$ & yes \\
        \texttt{VerbController} & $1978.0$ & $<0.001$ & $0.75$ & yes \\
        \texttt{PluralNoun} & $1928.5$ & $<0.001$ & $0.72$ & yes\\ 
        \texttt{NoTunnel} & $1696.0$ & $0.0093$ & $0.40$ & yes \\
        \texttt{RC415} & $1884.5$ & $0.0024$ & $0.28$ & yes \\
        \texttt{RC401} & $1756.0$ & $0.0093$ & $0.24$ & yes \\
        \texttt{PathHierarchy3} & $944.0$ & $0.9961$ & - & no \\
        \hline
        \hline
    \end{tabular}
\end{table}

\begin{figure}[H]
    \centering
    \includegraphics[width=\textwidth]{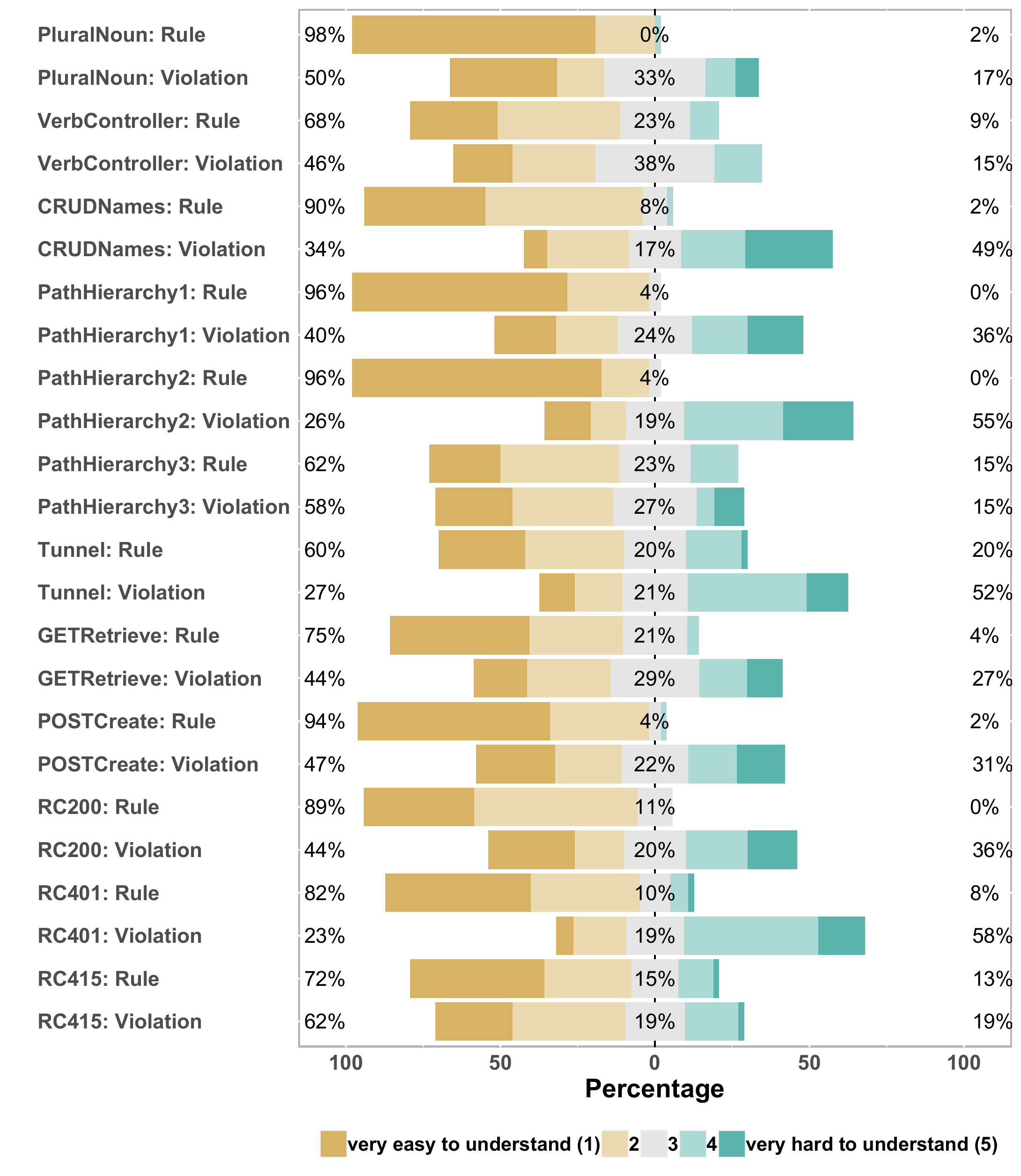}
    \caption{Likert plots of perceived difficulty ratings (RQ2)}
    \label{fig:likertPlots}
\end{figure}

\begin{table}[H]
	\caption{Hypothesis testing results for perceived difficulty ratings (RQ2), Holm-Bonferroni adjusted p-values, $\alpha = 0.05$, sorted by effect size}
	\label{tab:testResultsRQ2}
	\centering
	\begin{tabular}{l r r r r}
        Task & Test statistic (U) & p-value & Cohen's $d$ & Accepted \\
        \hline
        \hline
        \texttt{PathHierarchy2} & $2462.0$ & $<0.001$ & $2.06$ & yes \\
        \texttt{RC401} & $2297.5$ & $<0.001$ & $1.57$ & yes \\
        \texttt{PathHierarchy1} & $2203.0$ & $<0.001$ & $1.53$ & yes \\
        \texttt{CRUDNames} & $2254.5$ & $<0.001$ & $1.52$ & yes \\
        \texttt{POSTCreate} & $2076.0$ & $<0.001$ & $1.18$ & yes \\
        \texttt{PluralNoun} & $2069.5$ & $<0.001$ & $1.18$ & yes\\ 
        \texttt{NoRC200Error} & $1851.0$ & $<0.001$ & $0.94$ & yes \\
        \texttt{GETRetrieve} & $1972.5$ & $<0.001$ & $0.87$ & yes \\
        \texttt{NoTunnel} & $1839.5$ & $<0.001$ & $0.79$ & yes \\
        \texttt{VerbController} & $1676.0$ & $0.0701$ & - & no \\
        \texttt{RC415} & $1642.5$ & $0.0774$ & - & no \\
        \texttt{PathHierarchy3} & $1388.0$ & $0.4051$ & - & no \\
        \hline
        \hline
    \end{tabular}
\end{table}

\begin{table}[H]
    \caption{Correlation between TAU and perceived difficulty for the treatment \textit{violation}, sorted by correlation strength, Holm-Bonferroni adjusted p-values, $\alpha = 0.05$, insignificant correlations are marked with $\dagger$, difficulty ratings and TAU provided for comparison}
	\label{tab:correlationsRQ2}
	\centering
	\begin{tabular}{l r r | l l l l}
        \multirow{2}{*}{Task} &
        \multirow{2}{*}{Kendall's $\tau$} &
        \multirow{2}{*}{p-value} &
        \multicolumn{2}{c}{Median difficulty} &
        \multicolumn{2}{c}{Mean TAU} \\
        & & & \textit{rule} & \textit{violation} & \textit{rule} & \textit{violation} \\
        \hline
        \hline
        \texttt{NoTunnel} & 0.2264 & 1.0000 $\dagger$ & 2 & 4 & 0.4820 & 0.3396\\
        \texttt{CRUDNames} & 0.2124 & 1.0000 $\dagger$ & 2 & 3 & 0.7392 & 0.1982\\
        \texttt{PathHierarchy1} & -0.0574 & 1.0000 $\dagger$ & 1 & 3 & 0.8457 & 0.5678\\
        \texttt{POSTCreate} & -0.0634 & 1.0000 $\dagger$ & 1 & 3 & 0.8427& 0.4201\\
        \texttt{RC415} & -0.0950 & 0.9238 $\dagger$ & 2 & 2 & 0.7654 & 0.7044\\
        \texttt{NoRC200Error} & -0.1548 & 0.4996 $\dagger$ & 2 & 3 & 0.7268 & 0.3776\\
        \texttt{RC401} & -0.2516 & 0.0692 $\dagger$ & 2 & 4 & 0.6058 & 0.5170\\
        \texttt{PathHierarchy2} & -0.2582 & 0.0609 $\dagger$ & 1 & 4 & 0.7916 & 0.4913\\
        \texttt{PluralNoun} & -0.2595 & 0.0609 $\dagger$ & 1 & 2.5 & 0.8369 & 0.7046\\ 
        \texttt{PathHierarchy3} & -0.3435 & $0.0062$ & 2 & 2 & 0.6141 & 0.7505\\
        \texttt{VerbController} & -0.3543 & $0.0062$ & 2 & 3 & 0.7617 & 0.5345\\
        \texttt{GETRetrieve} & -0.5078 & $<0.001$ & 2 & 3 & 0.8095 & 0.4337\\
        \hline
        \hline
    \end{tabular}
\end{table}

\end{document}